\documentclass[english,11pt,aps,prd,a4paper,preprintnumbers,nofootinbib,showpacs,superscriptaddress,notitlepage]{revtex4}
\usepackage{graphicx}
\usepackage{amsmath}
\usepackage{amssymb}
\usepackage{hyperref}
\usepackage{xcolor}
\usepackage{ulem}
\usepackage{multirow}
\newcommand {\ignore}[1]{}
 \newcommand{\AddrUNAM}{Instituto de Ciencias Nucleares, Universidad Nacional Aut\'onoma de M\'exico, 04510 CDMX, M\'exico.}
 \newcommand{\AddrAHEP}{AHEP Group, Institut de F\'{i}sica Corpuscular --C.S.I.C./Universitat de Val\`{e}ncia, Parc Cientific de Paterna.\\C/Catedr\'atico Jos\'e Beltr\'an, 2 E-46980 Paterna (Val\`{e}ncia) - SPAIN}
\newcommand{\AddrCinvestav}{Departamento de F\'{\i}sica, Centro de
  Investigaci{\'o}n y de Estudios Avanzados del IPN\\ Apdo. Postal 14-740 07000 Ciudad de M\'exico, M\'exico}
\newcommand{\AddrTEC}{Tecnol\'ogico Nacional de M\'exico/ITS de Jerez,
  C.P. 99863 Zacatecas, M\'exico.}

\definecolor{darkred}{rgb}{0.6,0,0}
\definecolor{darkcyan}{RGB}{0, 111, 111}
\definecolor{dgreen}{rgb}{0,0.5,0}

\usepackage{stackengine}

\newcommand{\eps}{\epsilon}

\newcommand{\faser}{FASER$\nu$~}

\begin{document}

\title{Examining the sensitivity of FASER to Generalized Neutrino Interactions}

\author{F. J. Escrihuela}\email{franesfe@alumni.uv.es}\affiliation{\AddrAHEP}
\author{L. J. Flores}\email{ljflores@jerez.tecnm.mx}\affiliation{\AddrTEC}
\author{O. G. Miranda}\email{omar.miranda@cinvestav.mx}\affiliation{\AddrCinvestav}
\author{Javier Rend\'{o}n}\email{jesus.rendon@correo.nucleares.unam.mx }\affiliation{\AddrUNAM}
\author{R. S\'anchez-V\'elez}\email{ricsv05@icloud.com}\affiliation{\AddrCinvestav}

\begin{abstract}
{\noindent   We investigate the sensitivity of the FASER$\nu$ detector, a novel experimental setup at the LHC, to probe and constrain generalized neutrino interactions (GNI). Employing a comprehensive theoretical framework, we model the effects of generalized neutrino interactions on neutrino-nucleon deep inelastic scattering processes within the FASER$\nu$ detector. By considering all the neutrino channels produced at the LHC, we perform a statistical analysis to determine the sensitivity of FASER$\nu$ to constrain these interactions. Our results demonstrate that FASER$\nu$ can place stringent constraints on the GNI effective couplings. Additionally, we study the relation between GNI and a minimal Leptoquark model where the SM is augmented by a singlet Leptoquark with hypercharge $1/3$. We have found that the sensitivities for various combinations of the Leptoquark Yukawa couplings are approximately $\mathcal{O}(1)$, particularly when considering a Leptoquark mass in the TeV range.. }
\end{abstract}

\maketitle

\section{Introduction}
There are several reasons to extend the standard model of particle physics (SM), but maybe the most obvious one is the fact that neutrinos have mass. In order to extend the SM, one looks for different new physics (NP) scenarios. There are basically two different approaches to study NP, one is to study a very specific model with fixed degrees of freedom and symmetries or symmetry patterns, and the other is to study the problem using an effective field theory (EFT) framework. In this work we chose the second approach, which has the advantage of being more general and less restrictive. The EFT approach that we are implementing is known as generalized neutrino interactions, or GNI for short. The GNI formalism is in some sense a generalization of the non-standard neutrino interactions (NSI) formalism \cite{Ohlsson:2012kf,Miranda:2015dra,Farzan:2017xzy,Proceedings:2019qno}. While NSI consider only vector and axial vector couplings in the effective field theory, GNI extend the framework to also include scalar, pseudoscalar and tensor couplings, thus having the most general effective Lagrangian. The reason behind this extension in the SM neutrino interactions has its roots in several phenomena, for example, vector and axial vector modifications to the SM couplings can be introduced as a way to give an explanation to neutrino masses in some models~\cite{Schechter1980gr,Mohapatra1986bd,foot:1988aq,hirsch:2004he,Malinsky:2005bi,Grimus:2006nb}, such modifications can be studied in a model independent way within the NSI framework. On the other hand, tensor interactions can be realized if the neutrino can interact electromagnetically through a non-zero magnetic moment \cite{Giunti:2014ixa,Miranda:2019wdy}. Finally, scalar neutrino interactions could arise in several leptoquark models \cite{Buchmuller:1986zs,Crivellin:2019dwb,Gargalionis:2020xvt}. Many studies have been paying attention to scalar neutrino interactions, motivated by experimental data from solar neutrinos~\cite{Ge:2018uhz,Khan:2019jvr} and cosmological observations like BBN and CMB~\cite{Huang:2017egl,Forastieri:2019cuf,Escudero:2019gvw,Venzor:2020ova, Venzor:2022hql}.

Generalized neutrino interactions have been studied in a variety of neutrino experiments like neutrino-electron and neutrino-quark scattering, and also in beta decays \cite{Bischer:2018zcz,Bischer:2019ttk,Han:2020pff,Li:2020wxi,Chen:2021uuw,Escrihuela:2021mud}. Additionally, there are studies of GNI in coherent elastic neutrino-nucleus scattering~\cite{Lindner:2016wff}, using data from COHERENT~\cite{Papoulias:2017qdn,AristizabalSierra:2018eqm,Li:2020lba,Flores:2021kzl,DeRomeri:2022twg} and from the Dresden-II reactor~\cite{AristizabalSierra:2022axl,Majumdar:2022nby}. Projections of the sensitivity to GNI for future experiments such as PTOLEMY~\cite{Banerjee:2023lrk} and the DUNE near detector~\cite{Melas:2023olz} have also been made.

Our purpose here is to complement previous analyses by studying the sensitivity of the FASER$\nu$ experiment \cite{FASER:2019dxq} to generalized neutrino interactions. Being the first experiment to directly detect neutrinos from a particle collider~\cite{FASER:2023zcr}, \faser provides a great opportunity to study neutrinos and antineutrinos with energies of $10-1000$~GeV from all flavors due to its high flux. Moreover, it is also suitable to search for NP signals involving the neutrino sector, like heavy neutral leptons~\cite{Kling:2018wct, Ansarifard:2021elw}, sterile neutrinos~\cite{FASER:2019dxq}, charged and neutral current NSI~\cite{Ismail:2020yqc,Falkowski:2021bkq}, new vector mediators~\cite{Cheung:2021tmx}, nonunitary mixing~\cite{Aloni:2022ebm}, and neutrino electromagnetic properties~\cite{MammenAbraham:2023psg}.

This work is organized as follows. In section~\ref{sec:formalism}, we introduce the general formalism for the generalized neutrino interactions, including a detailed calculation of the relevant cross section for the FASER$\nu$ detector. In section~\ref{sec:experimental}, we give a brief experimental description of the FASER$\nu$ detector, as well as a comprehensive explanation about the adopted statistical analysis. Our results on the constraints of the GNI parameters for different scenarios are presented in section~\ref{sec:results}, and later, in section~\ref{sec:leptoquark}, are compared with a minimal leptoquark model. Finally, we present our conclusions in section~\ref{sec:conclusions}.

\section{Effective Lagrangian and cross-sections}
\label{sec:formalism}

In this work we follow the GNI formalism in our calculations. We work with an effective field theory with $4$-fermion interaction Lagrangians which, additionally to the Standard Model Lagrangian, includes scalar, vector, axial-vector, and tensor interactions. We will analyze both, charged-current and neutral-current neutrino processes at FASER. 

We start our discussion with the charged-current neutrino interactions which are described by the Lagrangian \cite{Falkowski:2021bkq}
\begin{equation}\label{LCC}
\begin{split}
\mathcal{L}^\mathrm{CC}_{eff}=-\frac{2V_{jk}}{v^{2}}&\Big[(1+\epsilon^{jk,L}_{\alpha\beta})(\bar{u}^{j}\gamma^{\mu}P_{L}d^{k})(\bar{\ell}_{\alpha}\gamma_{\mu}P_{L}\nu_{\beta})+\epsilon^{jk,R}_{\alpha\beta}(\bar{u}^{j}\gamma^{\mu}P_{R}d^{k})(\bar{\ell}_{\alpha}\gamma_{\mu}P_{L}\nu_{\beta})\\&+\frac{1}{2}(\epsilon^{jk,S}_{\alpha\beta})(\bar{u}^{j}d^{k})(\bar{\ell}_{\alpha}P_{L}\nu_{\beta})-\frac{1}{2}(\epsilon^{jk,P}_{\alpha\beta})(\bar{u}^{j}\gamma^{5}d^{k})(\bar{\ell}_{\alpha}P_{L}\nu_{\beta})\\&+\frac{1}{4}(\epsilon^{jk,T}_{\alpha\beta})(\bar{u}^{j}\sigma^{\mu\nu}P_{L}d^{k})(\bar{\ell}_{\alpha}\sigma_{\mu\nu}P_{L}\nu_{\beta})+h.c.\Big]\,,
\end{split}
\end{equation}
where $v=(\sqrt{2}G_{F})^{-1/2}\approx 246$ GeV with $G_{F}$ the Fermi constant, $P_{L,R}=\frac{1}{2}(1\mp\gamma^{5})$, $\sigma^{\mu\nu}=\frac{i}{2}[\gamma^{\mu},\gamma^{\nu}]$, and $V_{jk}$ the entries of the  Cabibbo-Kobayashi-Maskawa (CKM) matrix, and $\epsilon^{jk,X}_{\alpha\beta}$ are the parameters describing the strength of the corresponding  GNI ($X=L,R,S,P,T$).

For the case of the neutral-current neutrino processes we use the effective $4$-fermion interaction Lagrangian \cite{Bischer:2018zcz,Bischer:2019ttk,Han:2020pff}
\begin{equation}\label{equation_1}
	\mathcal{L}^\mathrm{NC}_{eff}=-\frac{G_{F}}{\sqrt{2}}\sum_{j}\epsilon^{f,j}_{\alpha\beta}(\bar{\nu}_{\alpha}\mathcal{O}_{j}\nu_{\beta})(\bar{f}\mathcal{O}^{'}_{j}f)\,,\\
\end{equation}
where again $G_{F}$ is the Fermi constant, $f$ is a fermion with a given flavor, and $\nu_{\alpha}$ represents a neutrino with flavor $\alpha$. The operators $\mathcal{O}_{j}$, and $\mathcal{O}^{'}_{j}$ characterize the generalized interactions describing the new physics, and the couplings $\epsilon^{f,j}_{\alpha\beta}$ give us information about the strength of these interactions. We show these operators and couplings in table ~\ref{tab:table_1}.

With these definitions for the Lagrangians, we can compute the cross sections that will be relevant for our analysis and will be discussed in the following subsections.

\begin{table}[!hbt]
	\centering
	\begin{tabular}{ccccc}
		\hline  \hline
		${\epsilon}$ & & $\mathcal{O}_{j}$ & &$\mathcal{O}^{'}_{j}$\\
		\hline \hline
		$\epsilon^{f,L}$ & &$\gamma_{\mu}(1-\gamma^{5})$ & &$\gamma^{\mu}(1-\gamma^{5})$\\
		\hline
		$\epsilon^{f,R}$ & &$\gamma_{\mu}(1-\gamma^{5})$ & &$\gamma^{\mu}(1+\gamma^{5)}$\\
		\hline
		$\epsilon^{f,S}$ & &$(1-\gamma^{5})$ & & $1$\\
		\hline
		$-\epsilon^{f,P}$ & &$(1-\gamma^{5})$ & &$\gamma^{5}$\\
		\hline
		$\epsilon^{f,T}$ & &$\sigma_{\mu\nu}(1-\gamma^{5})$ & &$\sigma^{\mu\nu}(1-\gamma^{5})$\\
		\hline \hline
	\end{tabular}
	\caption{Effective operators and effective couplings (see Eq.~\eqref{equation_1}) studied in this work.}
	\label{tab:table_1}
\end{table}

\subsection{Charged-current neutrino processes at \faser}

In this section we discuss processes of the form $\nu+N\rightarrow\ell+X$, where $\ell$ is a charged lepton, $N$ is a nucleon, and $X$ represents any final hadronic state.\\
\indent The differential cross-section for the charged-current interaction of a neutrino and a proton in a DIS process in the presence of GNI is given by
\begin{equation}\label{Cross_Section_CC}
\begin{split}
\frac{d\sigma^\mathrm{CC}_{\nu p}}{dxdQ^{2}} &= \frac{G^{2}_{\text{F}}}{8\pi} \Bigg[ 8 \left(1+\epsilon^{ud,L}_{\alpha\beta} \right)^{2}\Bigg( \left(1-\frac{m^{2}_{\ell}}{\hat{s}}\right)\sum_{q=d,s}q(x,Q^{2})+(\hat{s}-m^{2}_{\ell}-\hat{s}y)\left(\frac{1-y}{\hat{s}}\sum_{\bar{q}=\bar{u},\bar{c}}\bar{q}(x,Q^{2})\right) \Bigg)\\ &+ 8\left(\epsilon^{ud,R}_{\alpha\beta}\right)^{2}\Bigg(\left(\hat{s}-m^{2}_{\ell}-\hat{s}y\right)\left(\frac{1-y}{\hat{s}}\sum_{q=d,s}q(x,Q^{2})\right)+\left(1-\frac{m^{2}_{\ell}}{\hat{s}}\right)\sum_{\bar{q}=\bar{u},\bar{c}}\bar{q}(x,Q^{2})\Bigg)\\ &+ \left( \left(\epsilon^{ud,S}_{\alpha\beta}\right)^{2}+ \left(\epsilon^{ud,P}_{\alpha\beta}\right)^{2}\right)\left(y+\frac{m^{2}_{\ell}}{\hat{s}}\right)y\left(\sum_{q=d,s}q(x,Q^{2})+\sum_{\bar{q}=\bar{u},\bar{c}}\bar{q}(x,Q^{2})\right)\\ &+ 2\left(\epsilon^{ud,T}_{\alpha\beta}\right)^{2}\left(2 \left(1-\frac{m^{2}_{\ell}}{\hat{s}}\right)-y\left(y+\frac{m^{2}_{\ell}}{\hat{s}}\right)+2\left(\hat{s}-m^{2}_{\ell}-\hat{s}y\right)\left(\frac{1-y}{\hat{s}}\right)\Bigg)\left(\sum_{q=d,s}q(x,Q^{2})\right. \right. \\
+&  \sum_{\bar{q}=\bar{u},\bar{c}}\bar{q}(x,Q^{2})\Bigg)\Bigg]\,,
\end{split}
\end{equation}
where $\hat{s} \approx 2 m_N E_\nu x$ is the invariant mass squared of the neutrino-quark system, $m_\ell$ is the mass of the final state lepton, $y$ is the fraction of the neutrino momentum transferred to the hadronic state, $q(x,Q^{2})$ and $\bar{q}(x,Q^{2})$ are the parton distribution functions (PDFs) of the proton, that can be found in Ref. \cite{Martin:2009iq}. The arguments $x$ and $Q^2$ in the PDFs are, respectively, the Bjorken scaling variable and the (negative) squared four-momentum transfer, whose explicit forms are given by
\begin{equation}
x=\frac{Q^{2}}{2q\cdot p_{N}} \quad \text{and}\quad Q^{2}= -q^{2}\equiv-(p_{\nu}-p_{\ell})^{2} = \hat{s}y\,,
\end{equation}
where $p_{N}$ is the four-momentum of the nucleon, $p_{\nu}$ is the four-momentum of the neutrino, and $p_{\ell}$ is the four-momentum of the charged lepton.  \\
\indent  In Eq.\eqref{Cross_Section_CC} we have considered a universal GNI coupling $\eps_{\alpha\beta}^{ud,X}$ for simplicity, instead of a different coupling for each quark family (for a detailed study see e.g.~\cite{Falkowski:2021bkq}).  For the case of the SM we have: $\epsilon^{ud,L}_{\alpha\beta}=\epsilon^{ud,R}_{\alpha\beta}=\epsilon^{ud,S}_{\alpha\beta}=\epsilon^{ud,P}_{\alpha\beta}=\epsilon^{ud,T}_{\alpha\beta}=0$ so that Eq.~\eqref{Cross_Section_CC} reduces to \cite{Falkowski:2021bkq}

\begin{equation}
\frac{d\sigma^\mathrm{CC}_{\nu p}}{dxdQ^{2}}=\frac{G_F^2}{\pi}\Bigg[\left(1-\frac{m^{2}_{\ell}}{\hat{s}}\right)\sum_{q=d,s} q(x,Q^{2})+\left(\frac{\hat{s}-m^{2}_{\ell}-Q^{2}}{\hat{s}}\right)\left(1-\frac{Q^{2}}{\hat{s}}\right)\sum_{\bar{q}=\bar{u},\bar{c}}\bar{q}(x,Q^{2})\Bigg]\,.
\end{equation}
\indent For the differential cross section with the neutron $\frac{d\sigma^\mathrm{CC}_{\nu n}}{dxdQ^{2}}$, we just need to replace the corresponding parton distribution functions, that is, $q(x,Q^{2})\rightarrow q^{n}(x,Q^{2})$ and $\bar{q}(x,Q^{2})\rightarrow \bar{q}^{n}(x,Q^{2})$ with the following simplifications:
$u^{n}(x,Q^{2})=d(x,Q^{2})$ and $d^{n}(x,Q^{2})=u(x,Q^{2})$ by isospin symmetry  with analog relations for the corresponding antiquarks. For simplicity, we set the sea-quark contributions $s^{n}(x,Q^{2})=s(x,Q^{2})$ and $c^{n}(x,Q^{2})=c(x,Q^{2})$, with the same relations for the corresponding antiquarks as done in \cite{Falkowski:2021bkq}.\\

\begin{figure}[!hbt]
\includegraphics[scale=0.5]{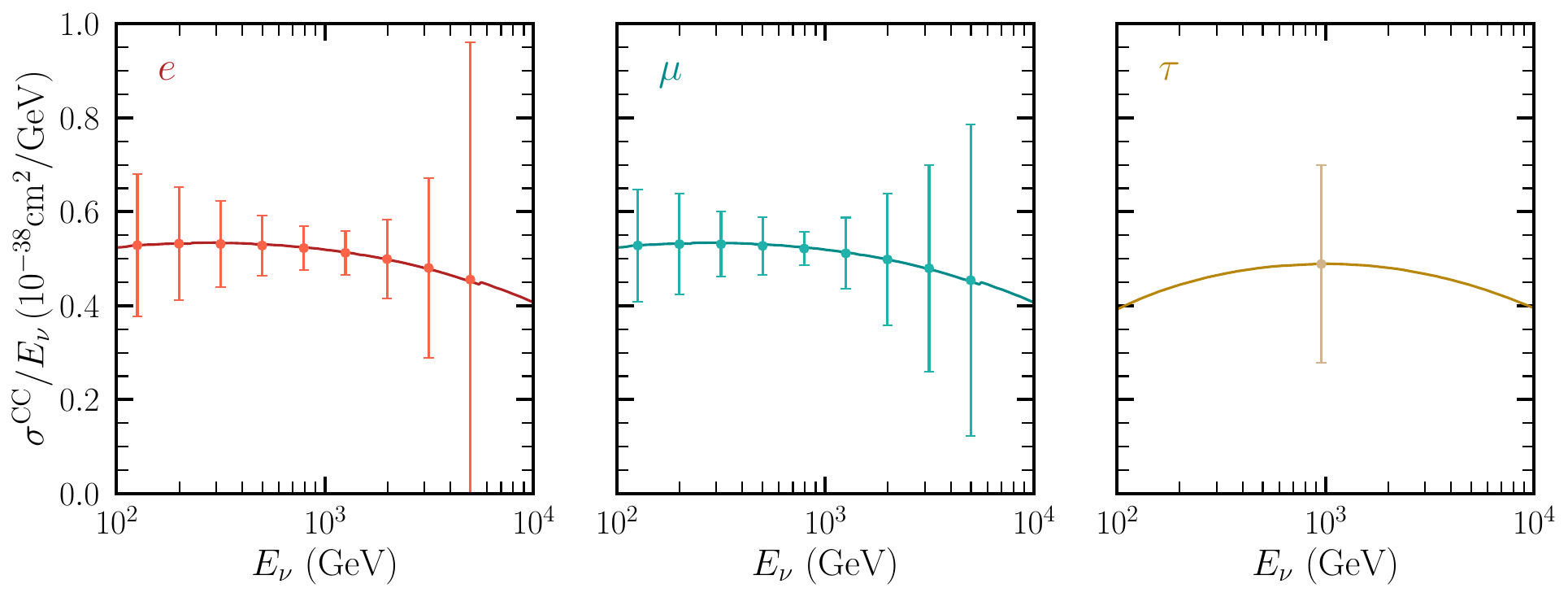}
\caption{Averaged neutrino and antineutrino charged-current cross section per tungsten nucleon as a function of incoming energy, for electron (left), muon (center), and tau (right) neutrinos. The solid lines are the SM expected cross sections, while the points represent the future \faser measurements along their uncertainties~\cite{FASER:2019dxq}.}
\label{fig:cc_crossSection}
\end{figure}

Note that charged currents always couple up quarks $(u,c)$ with down quarks $(d,s)$, thus, in these kind of processes we are not sensitive to individual couplings like $\epsilon^{u,X}_{\alpha\beta}$ or $\epsilon^{d,X}_{\alpha\beta}$ $(X=L, R, S, P, T)$, instead we are sensitive to one kind of parameter $\epsilon^{ud,X}_{\alpha\beta}$ that couples both types of quarks. For neutral currents we have a different situation, where up  and down couplings separate naturally.\\
The total cross section $\sigma^\mathrm{CC}_{\nu}$ is then easily found by summing up proton and neutron contributions in the nucleus, and integrating over the kinematic variables $x$ and $Q^{2}$, that is,
\begin{equation}
\sigma^{\text{CC}}_{\nu } = \int^1_{x_0}dx \int^{\hat{s}}_{Q^2_0} dQ^2 \left(n_n \frac{d^2 \sigma^{\text{CC}}_{\nu n}}{dx dQ^2} + n_p \frac{d^2 \sigma^{\text{CC}}_{\nu p}}{dx dQ^2}\right),
\end{equation}
with $n_{p}$ the number of protons, $n_{n}$ the number of neutrons, and with the lower limits of integration: $Q_0 \sim 1$ GeV and $x_0 = Q_0^2/(2 m_N E_\nu)$. For the case of \faser we have $n_n \simeq 110$ neutrons and $n_p =74$ protons. In Fig.~\ref{fig:cc_crossSection} we present our computations for  the SM predicted CC cross section per tungsten nucleon for each neutrino flavor, averaged over neutrino and antineutrino channels. The cross section sensitivity estimated by \faser for its full configuration is also shown in this figure. The expected uncertainties include statistical as well as systematic uncertainties coming from expected production rates.

\subsection{Neutral-current neutrino processes at \faser}

We turn now our attention to neutral-current deep inelastic neutrino scattering off a nucleon, which has the form $\nu+N\rightarrow\nu+X$ where, as in the charged-current case, $N$ represents any nucleon and $X$ is any final hadronic state. Unlike charged-current processes, the neutral-current process is significantly more difficult to identify. Whereas the CC process produces an outgoing lepton that carries much of the original neutrino energy, the NC neutrino interactions result in only a neutrino and any products of the recoiling nucleus. Despite its difficult measurement at the \faser experiment (which is primarily thought to identify CC events), neutral currents can also help us search for new physics signals, as it is shown in Ref. \cite{Ismail:2020yqc}.

\begin{figure}
\includegraphics[scale=0.6]{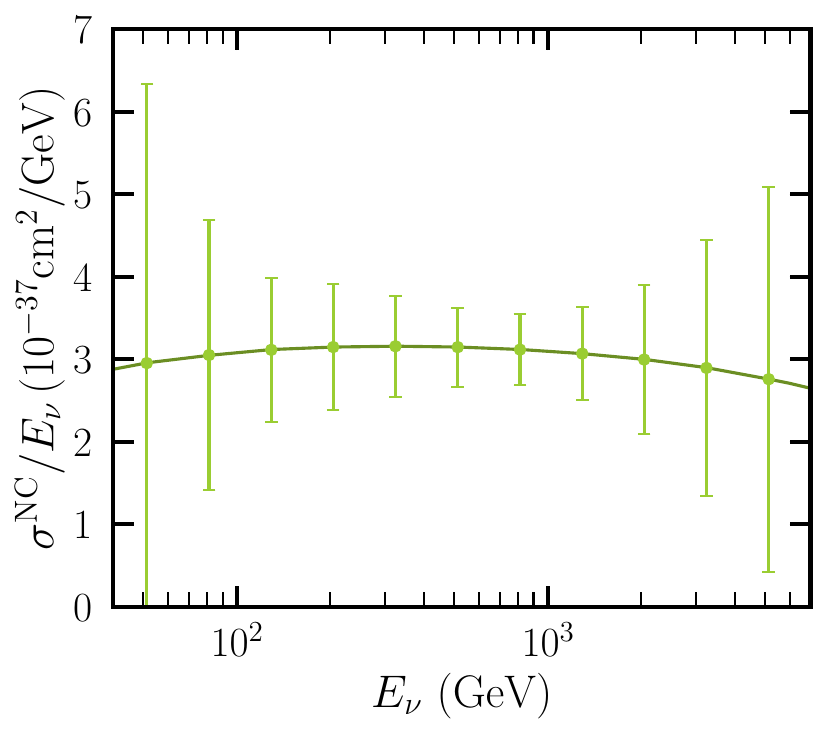}
\caption{Averaged neutrino and antineutrino neutral-current cross section per tungsten nucleus as a function of incoming energy. The solid line is the SM expected cross section, while the points represent the future \faser measurements along with its uncertainties~\cite{Ismail:2020yqc}.}
\label{fig:nc_crossSection}
\end{figure}

Using the Lagrangian in Eq.~\eqref{equation_1}, which induces the processes $\nu_{\alpha}(\bar{\nu}_{\alpha})+A \to \nu_{\beta}(\bar{\nu}_{\beta})+X$, where $A$ is the target nucleus and $X$ the final hadronic state, we can write the differential cross section for the neutrino-nucleon scattering process in the presence of the new general couplings as 

\begin{equation}\label{eq:crossSectionNC}
\begin{split}
\frac{d^2\sigma^\text{NC}_{\nu N}}{dx dQ^2} &= \frac{G^2_\text{F}}{8 \pi}\Bigg[ \sum_{q = u,c} q^N(x,Q^2) \Bigg( 8 \left(  \epsilon^{q,L}_\alpha \right)^2 + 8 (1-y)^2  \left	(\epsilon^{q,R}_\alpha \right)^2 + y^2 \left( (\epsilon^{q,P}_\alpha)^2+(\epsilon^{q,S}_\alpha)^2 \right)\\
& + 32 (2-y)^2 (\epsilon^{q,T}_\alpha)^2 \Bigg) + \sum_{\bar{q} = \bar{u},\bar{c}} \bar{q}^N(x,Q^2) \Bigg( 8 \left(\epsilon^{q,R}_\alpha \right)^2 + 8 (1-y)^2 \left(\epsilon^{q,L}_\alpha \right)^2\\
&+ y^2 \left((\epsilon^{q,P}_\alpha)^2 + (\epsilon^{q,S}_\alpha)^2\right) + 32 (2-y)^2 (\epsilon^{q,T}_\alpha)^2 \Bigg) + \Big(u \to d, c\to s \Big) \Bigg] ,
\end{split}
\end{equation}
where we have defined 
\begin{equation}
\begin{split}
\left(\epsilon^{q,X}_\alpha\right)^2 & = \sum_\beta \left(\epsilon^{q,X}_{\alpha\beta} + \delta_{\alpha\beta}\; g^{q,X}\frac{m^2_Z}{m^2_Z + \hat{s} y}\right)^2, \quad \text{for} \quad X = L,R, \\
\left(\epsilon^{q,X}_\alpha\right)^2 & = \sum_\beta (\epsilon^{q,X}_{\alpha\beta})^2, \quad \text{for} \quad X = P,S,T,
\end{split}
\end{equation}
with $g^{q,X}$ the SM couplings given by
\begin{equation}
g^{q,L} = T^q_3-Q_q \sin^2_W, \quad g^{q,R} = -Q_q \sin^2_W.
\end{equation}
In Eq.~\eqref{eq:crossSectionNC} we have ignored the quark masses and took into account the first and second  generation fermions only. Besides, we also consider the term of the $Z$ boson propagator (from the SM contribution), since at the \faser energies  we cannot neglect its contribution. Similar to the charged-current process, the neutral-current cross section is written in terms of the kinematic variable $Q^2$ and the quarks PDFs, which, again, we have taken the values from Ref.~\cite{Martin:2009iq}. However, unlike the charged-current scattering process, the contributions for the up quarks can be separated from the corresponding down quarks contribution. This feature allows us to explore the effective down- and up-quark couplings to neutrinos separately as well as under scenarios with identical coupling strengths. From Eq.~\eqref{eq:crossSectionNC} we can obtain the contribution for the antineutrino process by exchanging $q \leftrightarrow \bar{q}$ in the parton distribution functions. To obtain the total cross section for the neutrino scattering off nucleus, we proceed exactly as we did in the charged-current case. Therefore, we integrate over the kinematic variables $x$ and $Q^2$ and sum over the nucleons 
\begin{equation}
\sigma^{\text{NC}}_\nu  = \int^1_{x_0} dx \int^{\hat{s}}_{Q^2_0} dQ^2 \left(n_n \frac{d^2 \sigma^{\text{NC}}_{\nu n}}{dx dQ^2} + n_p \frac{d^2 \sigma^{\text{NC}}_{\nu p}}{dx dQ^2}\right) ,
\end{equation}
with a similar formula for the antineutrino cross section $\sigma^{\text{NC}}_{\bar{\nu}}$.
As we have pointed out previously, the lower limits of integration are: $Q_0 \sim 1$ GeV and $x_0 = Q_0^2/(2 m_N E_\nu)$. The NC differential cross section for neutrino scattering in the SM framework is obtained by neglecting all the GNI parameters in Eq.~\eqref{eq:crossSectionNC}, that is,
\begin{equation}\label{eq:crossSectionNC_SM}
\begin{split}
\frac{d^2\sigma^\text{NC(SM)}_{\nu N}}{dx dQ^2} &= \frac{G^2_\text{F}}{\pi} \frac{m^2_Z}{m^2_Z + \hat{s} y}\Bigg[ \sum_{q = u,c} q^N(x,Q^2) \left( \left(g^{u,L}\right)^2 + (1-y)^2  \left(g^{u,R}\right)^2 \right) \\
& + \sum_{\bar{q} = \bar{u},\bar{c}} \bar{q}^N(x,Q^2) \left( (1-y)^2 \left( g^{u,L} \right)^2 + \left( g^{u,R} \right)^2 \right) + \Big(u \to d, c\to s \Big) \Bigg] .
\end{split}
\end{equation}
In Fig.~\ref{fig:nc_crossSection} we show the theoretical prediction for the SM cross section averaged over neutrinos and antineutrinos per tungsten nucleon as a function of the incoming neutrino energy, $E_\nu$. In this plot, the dots show the expected \faser measurements and the bars correspond to their expected total uncertainties, taking into account uncertainties associated with background simulation and  neutrinos production ratios, according to Ref.~\cite{Ismail:2020yqc}. 

\section{Experimental description and analysis}
\label{sec:experimental} 

The \faser experiment provides a great opportunity to test new physics scenarios given the abundant neutrino flux produced at the LHC. By
identifying the charged lepton produced after the interaction with the detector, the \faser capacity to measure charged-current (CC)
interactions has been thoroughly studied by the FASER collaboration. With a total Tungsten target mass of 1.2 tons and a baseline of 480 m, the \faser detector will be able to measure approximately $\sim$1300 electron neutrinos, $\sim$20000 muon neutrinos and $\sim$20 tau neutrinos, with energies of $\mathcal{O}(100)$ GeV up to 1 TeV~\cite{FASER:2019dxq}. Moreover, the first 153 registered muon neutrino CC events have been recently reported by \faser~\cite{FASER:2023zcr},  proving that the experiment is feasible and will provide in the future precise measurements in these processes. To obtain a forecast of the sensitivity to GNI in the \faser charged-current future data, we will assume that the experiment will measure the SM prediction in each bin, $\sigma^{\text{CC(SM)}}_{i}$, with a standard deviation, $\sigma_i$, equal to the expected uncertainties to be achieved by \faser once the systematic are well understood. With these assumptions we will perform a $\chi^2$ analysis with the function
\begin{equation}
  \label{chi2}
\chi^2_{{\text{CC}_\alpha}} = \sum_{i} \left( \frac{\sigma^{\text{CC(SM)}}_{\alpha i}-\sigma^{\text{CC}}_{\alpha i} }{\sigma_{\alpha i}} \right)^2
\end{equation}
where $\alpha$ stands for the corresponding flavor that we analyze, and $i$ runs for the number of bins in the given flavor sample of Fig.~\ref{fig:cc_crossSection}. We will assume that $\sigma^{\text{CC}}_{i}$ depends on the GNI parameter under study and will get the corresponding sensitivity region.

On the other hand, the measurement of neutral-current (NC) interactions is more challenging, given the absence of charged leptons in the final state and the possible contamination from misidentified CC events and neutral hadron interactions. Nevertheless, a recent study argued that NC events can be identified using a neural network-based analysis~\cite{Ismail:2020yqc}. In this work, we also consider this possibility, and the expected sensitivity for the GNI couplings are obtained by using a $\chi^2$ function analog to the one used for charged currents in Eq.~\eqref{chi2}, replacing the charged-current theoretical prediction for the SM and GNI, by the corresponding neutral-current expectation: 
\begin{equation}
	\label{eq:chi2_NC}
	\chi^2_\mathrm{NC} = \sum_{i} \left( \frac{\sigma^{\text{NC(SM)}}_{i}-\sigma^{\text{NC}}_{i} }{\sigma_{i}} \right)^2.
\end{equation}
In this neutral-current case we use the corresponding \faser expected uncertainties shown in Fig.~\ref{fig:nc_crossSection}

\ignore{
\begin{equation}
\chi^2_{\text{NC}} = \sum_i \left( \frac{\sigma^{\text{NC(SM)}}_i-\sigma^{\text{NC}}_i }{\sigma_i} \right)^2
\end{equation}
}

\section{Results}
\label{sec:results}
\begin{figure}[!hbt]
\includegraphics[scale=0.6]{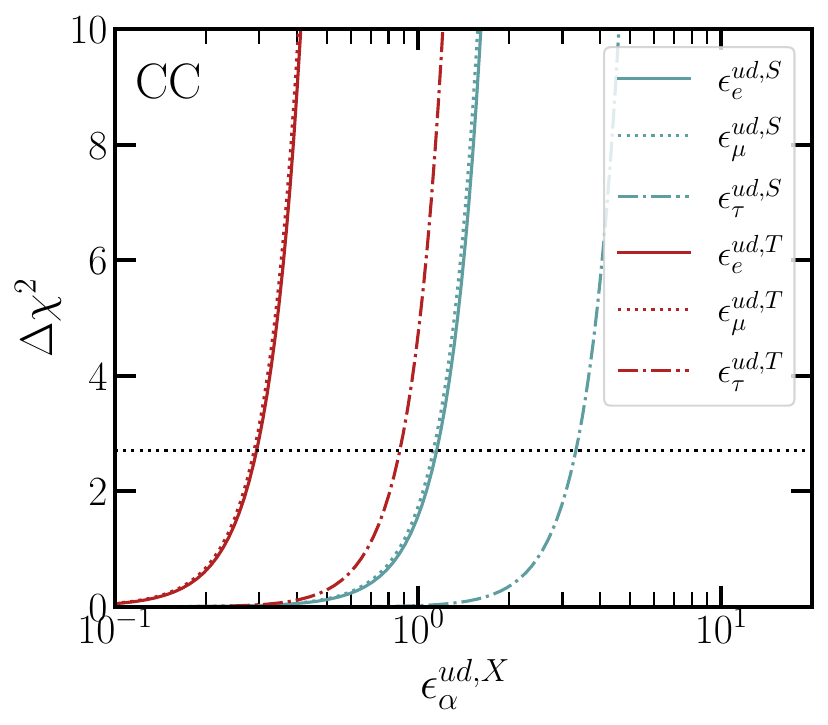}
\caption{$\chi^2$ profiles of the \faser expected sensitivity to scalar (blue) and tensor (red) charged-current GNI parameters.}
\label{fig:cc_scalar_tensor_1D}
\end{figure}

\begin{figure}[!hbt]
\includegraphics[scale=0.38]{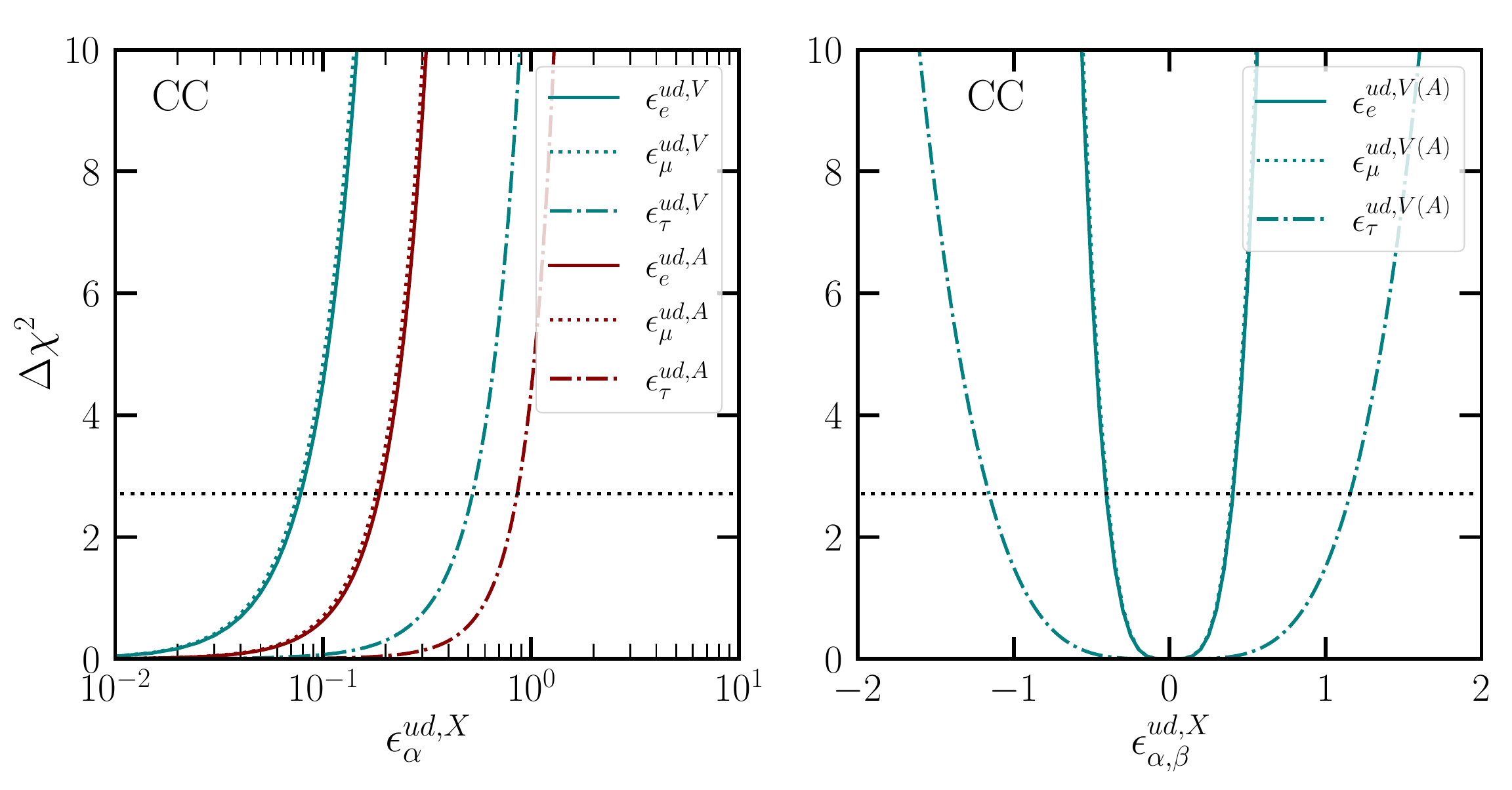}
\caption{$\chi^2$ profiles of the \faser expected sensitivity to vector (blue) and axial (red) charged-current GNI diagonal (left panel) and non-diagonal (right panel) parameters.}
\label{fig:cc_vector_axial_1D}
\end{figure}

\begin{figure}[!hbt]
\includegraphics[scale=0.6]{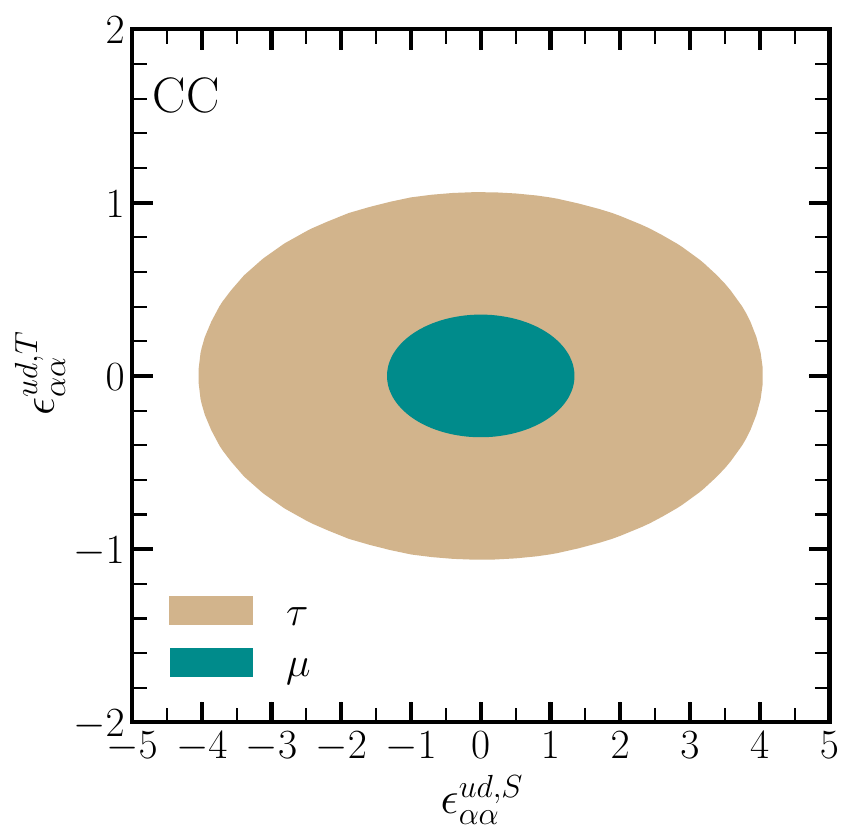}
\caption{Expected allowed regions at 95\% C.L. for the scalar and tensor charged-current GNI parameters, from the muon- and tau-neutrino beams. }
\label{fig:cc_scalar_tensor_2D}
\end{figure}

\begin{figure}[!hbt]
\includegraphics[scale=0.5]{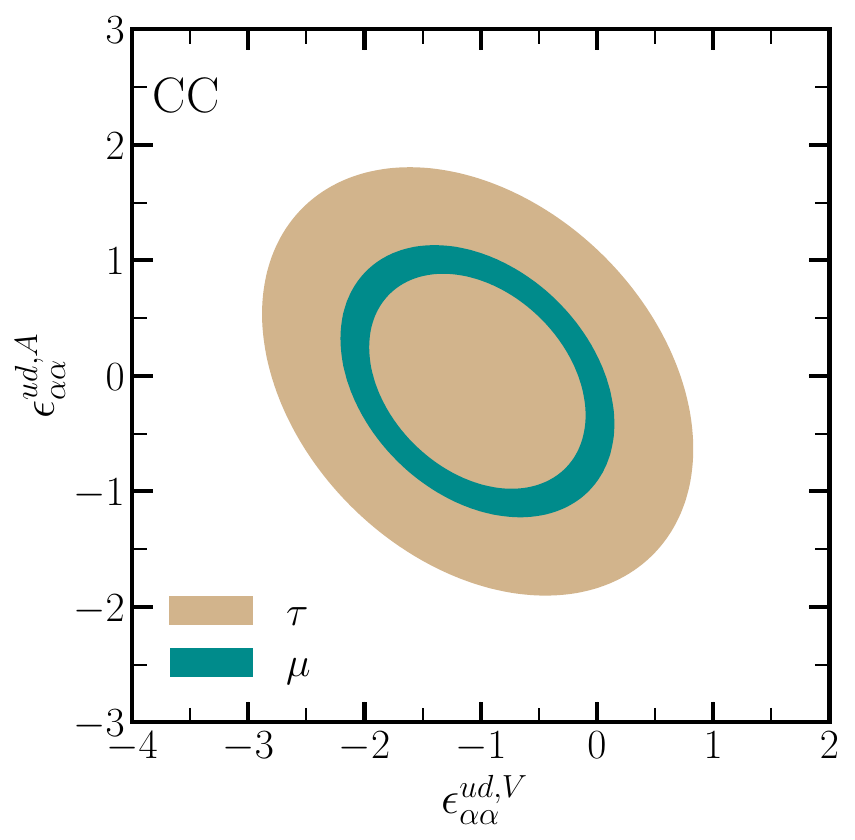}
\includegraphics[scale=0.5]{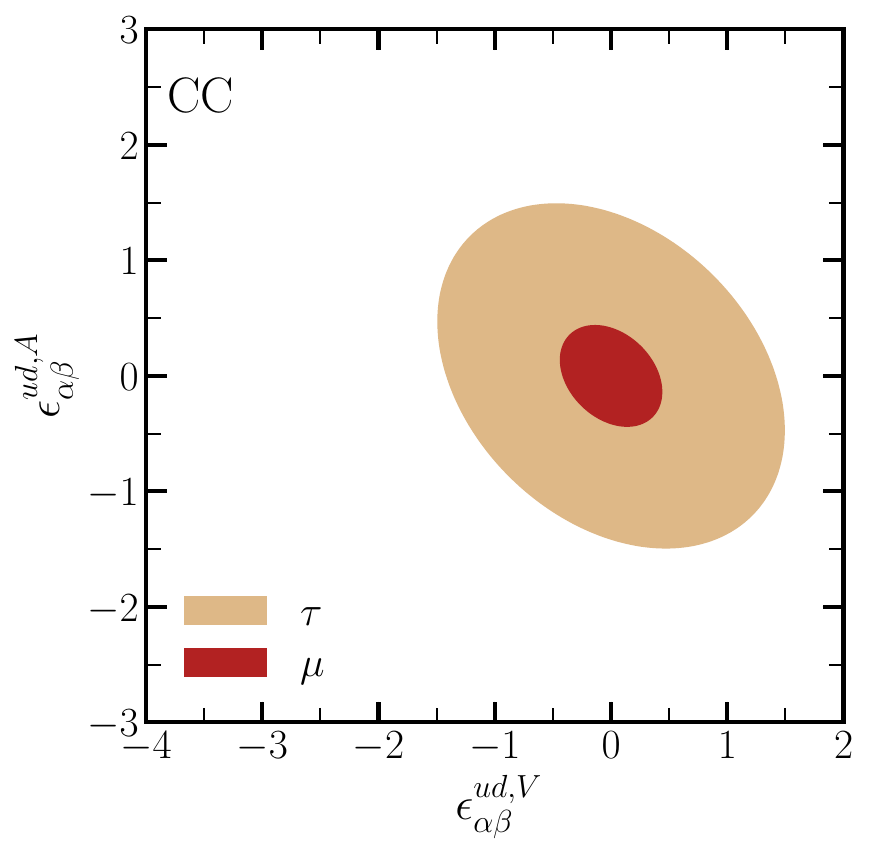}
\caption{Expected allowed regions at 95\% C.L. for the vector and axial-vector charged-current GNI parameters, from the muon- and tau-neutrino beams. The left (right) panel corresponds to the flavor-conserving (-changing) parameters. }
\label{fig:cc_vector_axial_2D}
\end{figure}

We have studied the expected sensitivity to the GNI parameters from future \faser results. We start our analysis by considering only one non-zero observable at a time. Our results for the charged-current case are shown in Figs.~\ref{fig:cc_scalar_tensor_1D}, ~\ref{fig:cc_vector_axial_1D}, ~\ref{fig:cc_scalar_tensor_2D}, and \ref{fig:cc_vector_axial_2D}. In Fig.~\ref{fig:cc_scalar_tensor_1D}, we see the $\chi^2$ profiles for scalar and tensor couplings. Pseudoscalar and scalar couplings have the same behavior as can be seen in Eq.~\eqref{Cross_Section_CC}, thus, the results from Fig.~\ref{fig:cc_scalar_tensor_1D} are valid for both couplings.

 Similarly, in Fig.~\ref{fig:cc_vector_axial_1D}, we see the $\chi^2$ profiles for vector and axial couplings. Unlike in the scalar or tensor cases, here we can separate diagonal and non-diagonal contributions due to the interference between the SM and the GNI. In the left and right panels we show the diagonal and non-diagonal bounds, respectively.  Muon flavor parameters have the strongest constraints, whereas tau flavor parameters have the weakest. This is easily understood in terms of statistics since muon neutrinos dominate the process allowing us to set better bounds, contrary to what happens with tau neutrinos where we have fewer events. This is a common feature in all our plots for the CC processes, and not just a particularity of Figs.~\ref{fig:cc_scalar_tensor_1D} and \ref{fig:cc_vector_axial_1D}. Another thing that we see from Fig.~\ref{fig:cc_scalar_tensor_1D} is that tensor interactions are better constrained than scalar interactions.

We continue our analysis by studying two GNI parameters simultaneously while we set the other parameters to zero. These results are shown in Fig.~\ref{fig:cc_scalar_tensor_2D} for the case of scalar and tensor interactions, where we show the allowed region for the scalar vs tensor plane $(\epsilon^{ud,S}_{\alpha\alpha}\,\, \text{vs}\,\, \epsilon^{ud,T}_{\alpha\alpha})$, and in Fig.~\ref{fig:cc_vector_axial_2D} for the case of vector and axial interactions, where we show the allowed region for the vector vs axial plane $(\epsilon^{ud,V}_{\alpha\alpha}\,\, \text{vs}\,\, \epsilon^{ud,A}_{\alpha\alpha})$. As we mentioned above, when we discuss vector or axial couplings we need to consider diagonal and non-diagonal cases, this is why we separate Fig.~\ref{fig:cc_vector_axial_2D} in two different panels. In the left panel we show the diagonal case and in the right panel we show the non-diagonal case. \\

\begin{figure}[th]
\includegraphics[scale=0.63]{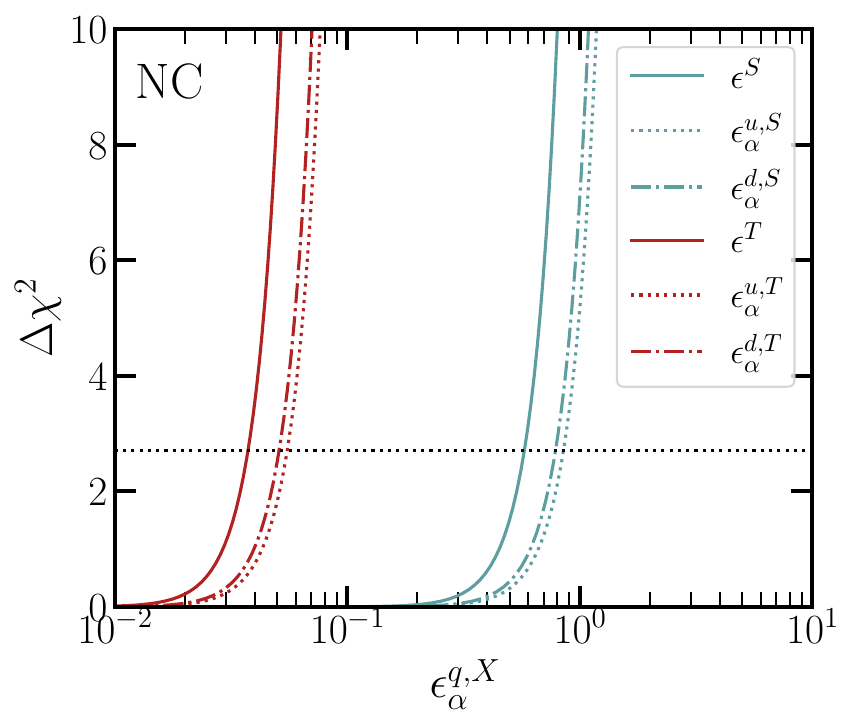}
\caption{$\chi^2$ profiles of the \faser expected sensitivity to scalar (blue) and tensor (red) neutral-current GNI parameters. The solid lines stand for the case where $\eps^{uX}=\eps^{dX}$, for both scalar and tensor couplings.}
\label{fig:nc_scalar_tensor_1D}
\end{figure}

We now turn our attention to the neutral-current process, where similar to the CC analysis, we first consider one non-zero GNI parameter at a time, following Eq.~\eqref{eq:chi2_NC} and using the appropriate errors from Fig.~\ref{fig:nc_crossSection}. In order to simplify the analysis, we assume $\epsilon^{c,X}_{\alpha \beta} = \epsilon^{u,X}_{\alpha \beta}$ and $\epsilon^{s,X}_{\alpha \beta} = \epsilon^{d,X}_{\alpha \beta}$, i.e quarks of the same family have equal strength. Under this assumption, the $\chi^2$ profiles for the scalar (blue lines) and tensor (red lines) GNI parameters are shown in Fig.~\ref{fig:nc_scalar_tensor_1D} for two scenarios: when up-quark couplings $\epsilon^{u,X}_{\alpha}$ and down-quark couplings $\epsilon^{d,X}_{\alpha}$ can be distinguished, and when they have equal strength $\epsilon^{u,X}_{\alpha} = \epsilon^{d,X}_{\alpha} \equiv \epsilon^{X}_{\alpha}$. Because of the symmetry in the NC cross section~\eqref{eq:crossSectionNC}, scalar and pseudoscalar couplings have equal strength, which is why the pseudoscalar profiles are not shown. As can be seen from this figure, \faser will be more sensitive to the tensor interaction when compared to the scalar and pseudoscalar interactions. This result is easy to understand because the tensor interaction is $32$ times stronger.

\begin{figure}[!hbt]
\includegraphics[scale=0.38]{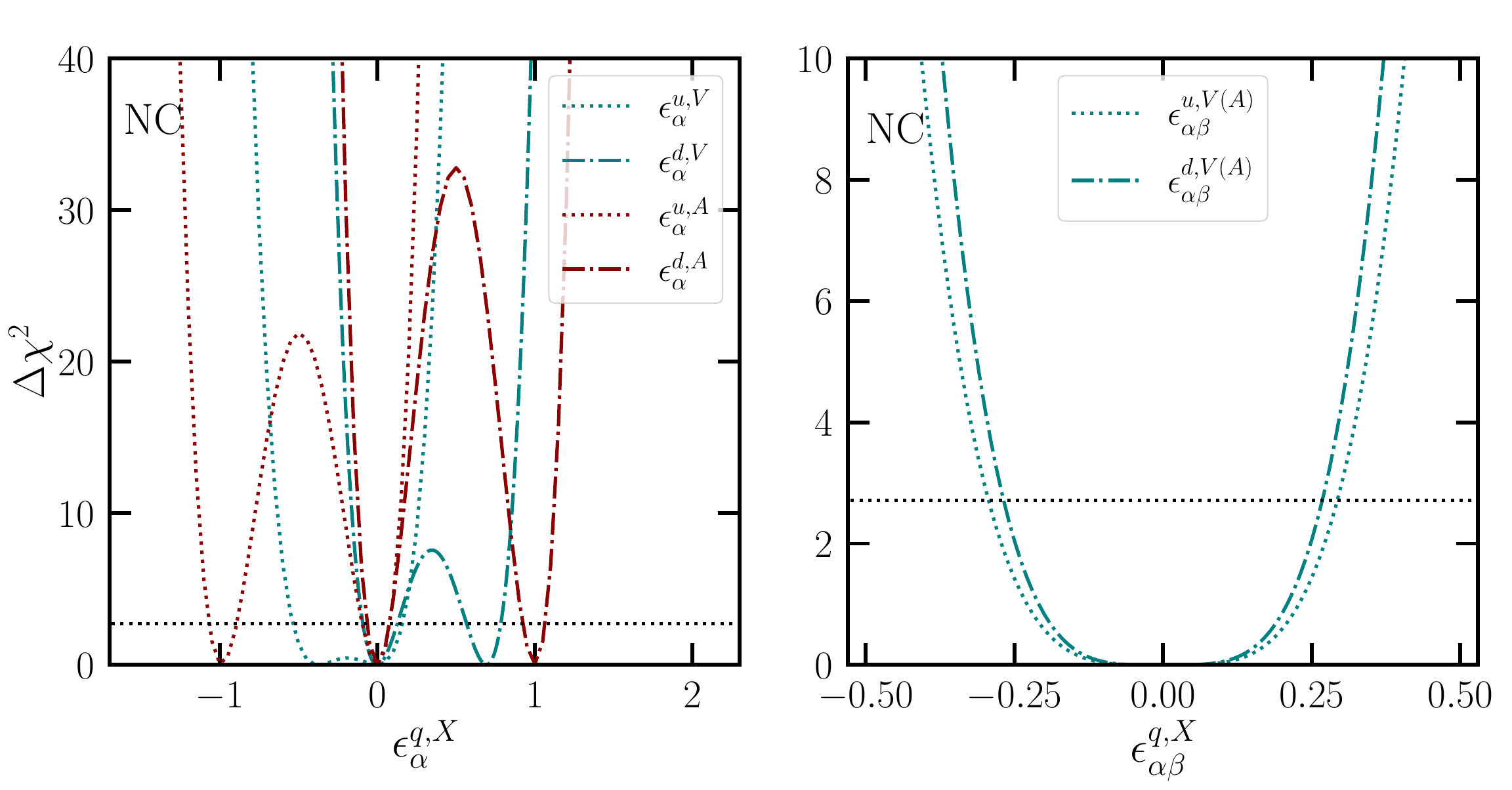}
\caption{$\chi^2$ profiles of the \faser expected sensitivity to vector (blue) and axial (red) neutral-current GNI parameters. The flavor-conserving(-changing) couplings are shown in the left (right) panel. }
\label{fig:nc_vector_axial_1D}
\end{figure}

Regarding the remaining GNI couplings, the left and right panels of Fig. \ref{fig:nc_vector_axial_1D} display the $\chi^2$ profiles for vector and axial GNI couplings for diagonal interactions and non-diagonal ones, respectively. In this scenario, we employed the standard definitions, provided by 
\begin{equation}
\epsilon^{q,V} = \epsilon^{q,L}+\epsilon^{q,R}\quad \text{and} \quad 
\epsilon^{q,A} = \epsilon^{q,L}-\epsilon^{q,R}.
\end{equation}   

As observed in the left panel of Fig. \ref{fig:nc_vector_axial_1D}, the new vector and axial couplings, when the down and up quarks interactions are different, the $\chi^2$ profiles have two minima. For the vector couplings, the minimal values are obtained in $\epsilon^{q,V}_\alpha = 0$, and the other one in $0.7$ for $\epsilon^{d,V}_\alpha$ and $-0.4$ for $\epsilon^{u,V}_\alpha$. Since the vector and axial GNI couplings interfere with the SM contribution, in the right panel we show the non-diagonal couplings, and in this case it turns out that both couplings have the same profile, with the minimum value at $\epsilon^{q,X}_{\alpha \beta} = 0$. From this NC analysis, we can see that the sensitivity for the up and down quarks couplings are very similar, since their PDF functions have no significant difference. However, this condition does not apply for the vector and axial diagonal couplings due to the SM contribution. The result for the case when we consider universal couplings to quarks $\epsilon^{u,X}_\alpha = \epsilon^{d,X}_\alpha$ is natural, since the improvement in  the sensitivity will be driven by the increase in the statistics by considering both quark types. In this case, although we consider that all neutrino flavors  can contribute to the signal, most of contribution will come from muon neutrinos and we can assume that the constrained GNI parameter is mainly for $ \epsilon^{qX}_\mu$. For each of the previously stated scenarios, the corresponding sensitivities at $90\%$ C.L. from the CC and NC 1-dimensional analyses are shown in Table ~\ref{tab:limits}.\\

\begin{figure}[!hbt]
\includegraphics[scale=0.55]{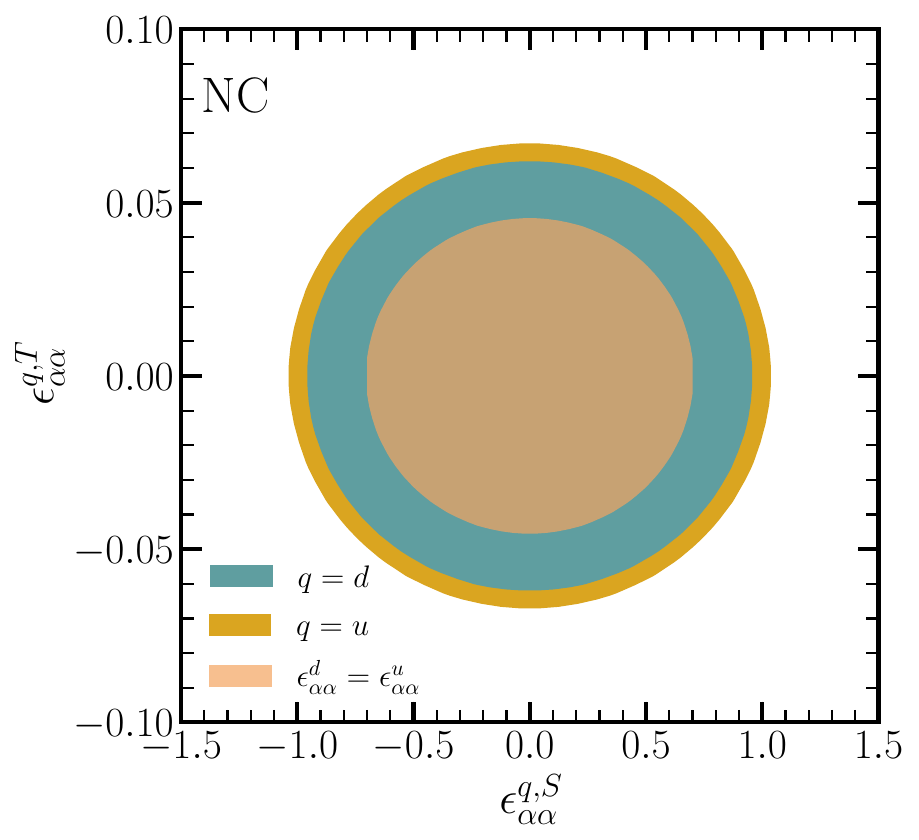}
\caption{Expected allowed regions at 95\% C.L. for the scalar and tensor neutral-current GNI parameters, from the muon- and tau-neutrino beams. }
\label{fig:nc_scalar_tensor_2D}
\end{figure}

We now carry out the analysis considering two free GNI parameters at the same time and set the rest of parameters to zero. Starting with the scalar and tensor interactions, we show in Fig.~(\ref{fig:nc_scalar_tensor_2D}) the allowed regions at $95\%$ C.L. in the S-T plane ($\epsilon^{q,S}_{\alpha\alpha}$ vs $\epsilon^{q,T}_{\alpha \alpha}$). We also show in this figure the case when we only consider the d-quark type contribution, the u-quark type, and the more restrictive case $\epsilon^{uX}_{\alpha} = \epsilon^{dX }_{\alpha}$. As observed, the allowed regions are centered around the SM solution $\epsilon^{q,S} = \epsilon^{q,T} = 0$. 

\begin{figure}[!hbt]
\includegraphics[scale=0.53]{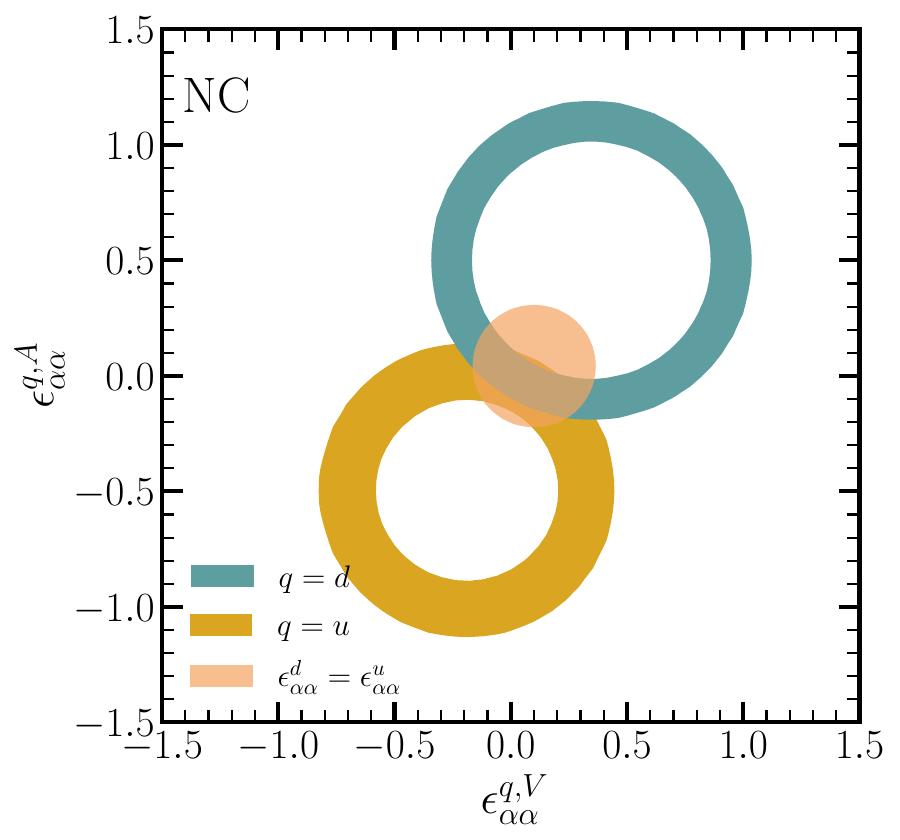}
\includegraphics[scale=0.53]{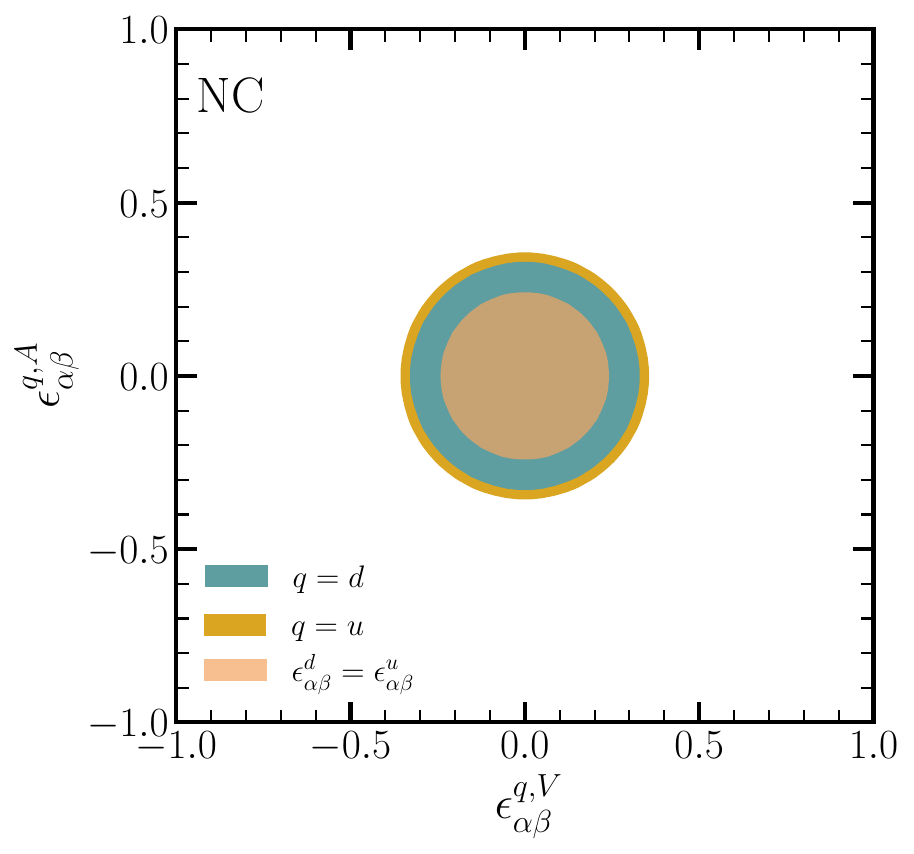}
\caption{Expected allowed regions at 95\% C.L. for the vector and axial-vector neutral-current GNI parameters. The left (right) panel corresponds to the flavor-conserving (-changing) parameters. Three scenarios are considered in both panels: only non-zero coupling to $d$ quarks (blue), only non-zero coupling to $u$ quarks (yellow), and equal coupling to $d$ and $u$ quarks (light-orange). }
\label{fig:nc_vector_axial_2D}
\end{figure}

In Fig.~\ref{fig:nc_vector_axial_2D} we show the allowed areas in the V-A plane for the cases of flavor-preserving (left plot) and flavor-changing (right plot) terms. It is worth mentioning that a similar result has been obtained in Ref.~\cite{Ismail:2020yqc}, however, the authors used a different convention for the axial GNI coupling, namely $\epsilon^{q,A}= \epsilon^{q,R}-\epsilon^{q,L}$. In the right panel, the scenario for non-diagonal GNI couplings is shown, where, as already noted from the one dimensional analysis, the allowed areas are centered in the SM solution $\epsilon^{q,V}_{\alpha\beta} = \epsilon^{q,A}_{\alpha\beta} = 0$.

\begin{figure}[!hbt]
\includegraphics[scale=0.55]{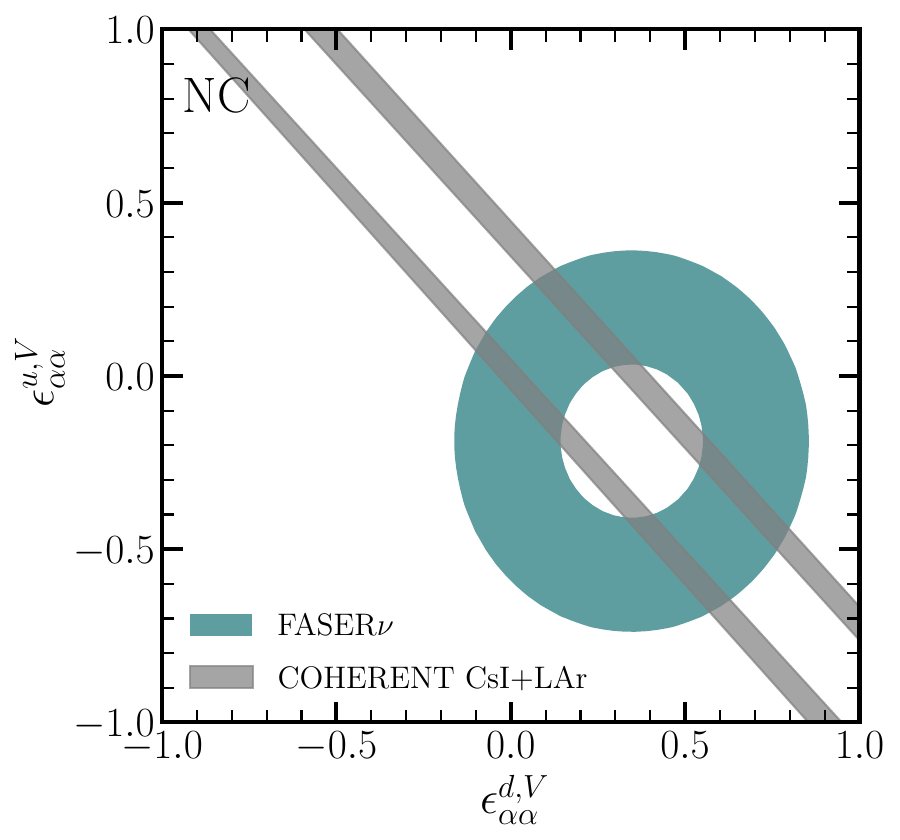}
\caption{Expected allowed region at 95\% C.L. for the vector neutral-current GNI coupling to $d$ and $u$ quarks. This region was obtained considering all three neutrino flavors, although the muon neutrino flux is significantly higher. The gray bands correspond to the current combined limit from the COHERENT CsI and LAr measurements~\cite{DeRomeri:2022twg}. }
\label{fig:nc_vector_2D}
\end{figure}

Finally, in Fig.~\ref{fig:nc_vector_2D}, the expected allowed area in the plane $\epsilon^{d,V}_{\alpha \alpha}$ vs $\epsilon^{u,V}_{\alpha \alpha}$ is depicted by considering all the neutrino flavors, although the muon neutrino flux is substantially higher than the flux of electron and tau neutrinos. In addition, we show the result from CE$\nu$NS when analyzing CsI and LAr COHERENT data~\cite{DeRomeri:2022twg}, where the combine analysis lead to the gray narrow bands depicted in the figure. Then, our result using the \faser data, is complementary in order to restrict the sensitivity to the expected region of the new vector couplings.

\begin{table}[!hbt]
	\centering
	\begin{tabular}{c@{\hskip 0.2in}c|c@{\hskip 0.2in}c} \hline 
		\toprule
		 \multicolumn{2}{c}{Charged current (CC)} & \multicolumn{2}{c}{Neutral current (NC)}  \\
		 \toprule
		 Parameters & 90\% C.L. limit & Parameters & 90\% C.L. limit\\   \hline  \hline
		$|\eps^{ud,V}_{ee}|$ &  $< 0.078$  &  $\eps^{u,V}_{\alpha\alpha}$ & $(-0.535,0.158)$ \\ 
		$|\eps^{ud,V}_{\mu\mu}|$ &  $<0.076$   & $\eps^{d,V}_{\alpha\alpha}$ &  $(-0.091, 0.126) \cup (0.566, 0.786)$ \\
		$|\eps^{ud,,V}_{\tau\tau}|$ &  $<0.526$  & $|\eps^{V}_{\alpha\alpha}|$ &  $(-0.119, 0.326)$ \\  
		$|\eps^{ud,V}_{e\beta}|$ &  $<0.404$  &  $|\eps^{u,V}_{\alpha\beta}|$ & $<0.293$ \\  
		$|\eps^{ud,V}_{\mu \beta}|$ &  $<0.401$  & $|\eps^{d,V}_{\alpha\beta}|$ & $<0.267$ \\ 
		$|\eps^{ud,V}_{\tau \beta}|$ &  $<1.156$ & $|\eps^{V}_{\alpha\beta}|$ & $<0.197$   \\  \hline
		$|\eps^{ud,A}_{ee}|$ &  $<0.188$  & $\eps^{u,A}_{\alpha\alpha}$ &  $(-1.06,-0.893)\cup(-0.094,0.075)$\\ 
		$|\eps^{ud,A}_{\mu\mu}|$ &  $<0.182$  & $\eps^{d,A}_{\alpha\alpha}$ &  $(-0.062, 0.075)\cup(0.924, 1.063)$\\
		$|\eps^{ud,A}_{\tau\tau}|$ &  $<0.865$  & $|\eps^{A}_{\alpha\alpha}|$ &   $(-0.155,	0.250)$\\  
		$|\eps^{ud,A}_{e\beta}|$ &  $<0.404$  & $|\eps^{u,A}_{\alpha\beta}|$ &  $<0.293$\\  
		$|\eps^{ud,A}_{\mu \beta}|$ &  $<0.401$  & $|\eps^{d,A}_{\alpha\beta}|$ &  $<0.267$\\ 
		$|\eps^{ud,A}_{\tau \beta}|$ &  $<1.156$  & $|\eps^{A}_{\alpha\beta}|$ &  $< 0.197$\\  \hline
		$|\eps^{ud,S}_{e}|$ &  $<1.12$  & $|\eps^{u,S}|$ &  $<0.854$ \\ 
		$|\eps^{ud,S}_{\mu}|$ &  $<1.15$  & $|\eps^{d,S}|$ &  $<0.783$\\  
		$|\eps^{ud,S}_{\tau}|$ &  $<3.31$  & $|\eps^{S}|$ &  $<0.583$ \\  \hline
		$|\eps^{ud,T}_{e}|$ &   $<0.295$ & $|\eps^{u,T}|$ &   $<0.055$\\
		$|\eps^{ud,T}_{\mu}|$ &   $<0.291$  & $|\eps^{d,T}|$ &   $<0.051$\\
		$|\eps^{ud,T}_{\tau}|$ &   $<0.875$  & $|\eps^{T}|$ &   $<0.037$\\  \hline\hline
	\end{tabular} 
	\caption{\footnotesize  Expected 90\% C.L. limits for the charged and neutral current GNI parameters. The following definitions have been used throughout the table: $|\eps^{q,X}_\alpha|^2 = \sum_{\beta}  |\eps^{q,X}_{\alpha\beta}|^2$, $|\eps^{q,X}|^2 = \sum_{\alpha\beta}  |\eps^{q,X}_{\alpha\beta}|^2$.} \label{tab:limits}   
\end{table}

\section{Comparison with Leptoquark models}
\label{sec:leptoquark}

Different well motivated theories predict the existence of Leptoquarks (LQs) as scalar or vector mediators that couple a quark-lepton pair at tree level. They may arise naturally in grand unified models~\cite{Georgi:1974sy,Ramond:1976jg,Senjanovic:1982ex}, as well as other well motivated theories such as Thechnicolor~\cite{Hill:2002ap}, models with composite fermions~\cite{Schrempp:1984nj,Gripaios:2009dq} or models with extended scalar sectors~\cite{Davies:1990sc}, for instance. However, it is more feasible to study the LQ phenomenology in a model-independent framework through an effective Lagrangian. The most general effective interaction invariant under the group $SU(3)_c\times SU(2)_L \times U(1)_Y$, for both scalar and vector LQs, was initially proposed in Ref.~\cite{Davidson:1993qk} and later analyzed in Ref.~\cite{Dorsner:2016wpm}. In this work, we study the scalar LQ with quantum numbers $(\bar{\textbf{3}}, \textbf{1}, 1/3)$, i.e., it transforms as a singlet under $SU(2)$, usually denoted as $S_1$ in the literature. The phenomenology of LQs has been widely studied since it offers a possible solution for the anomalies in the semi-leptonic $B$ decays~\cite{Angelescu:2018tyl,Lee:2021jdr} and the muon anomalous magnetic moment~\cite{FileviezPerez:2021lkq,Gherardi:2020qhc,Dorsner:2019itg}. As a consequence, different studies have been carried out to constrain the LQ couplings through experimental data searching for LQs at the LHC and the IceCube neutrino experiment~\cite{Dorsner:2019vgp}. 

\begin{figure}[hbt!]
\centering
\includegraphics[width=10cm]{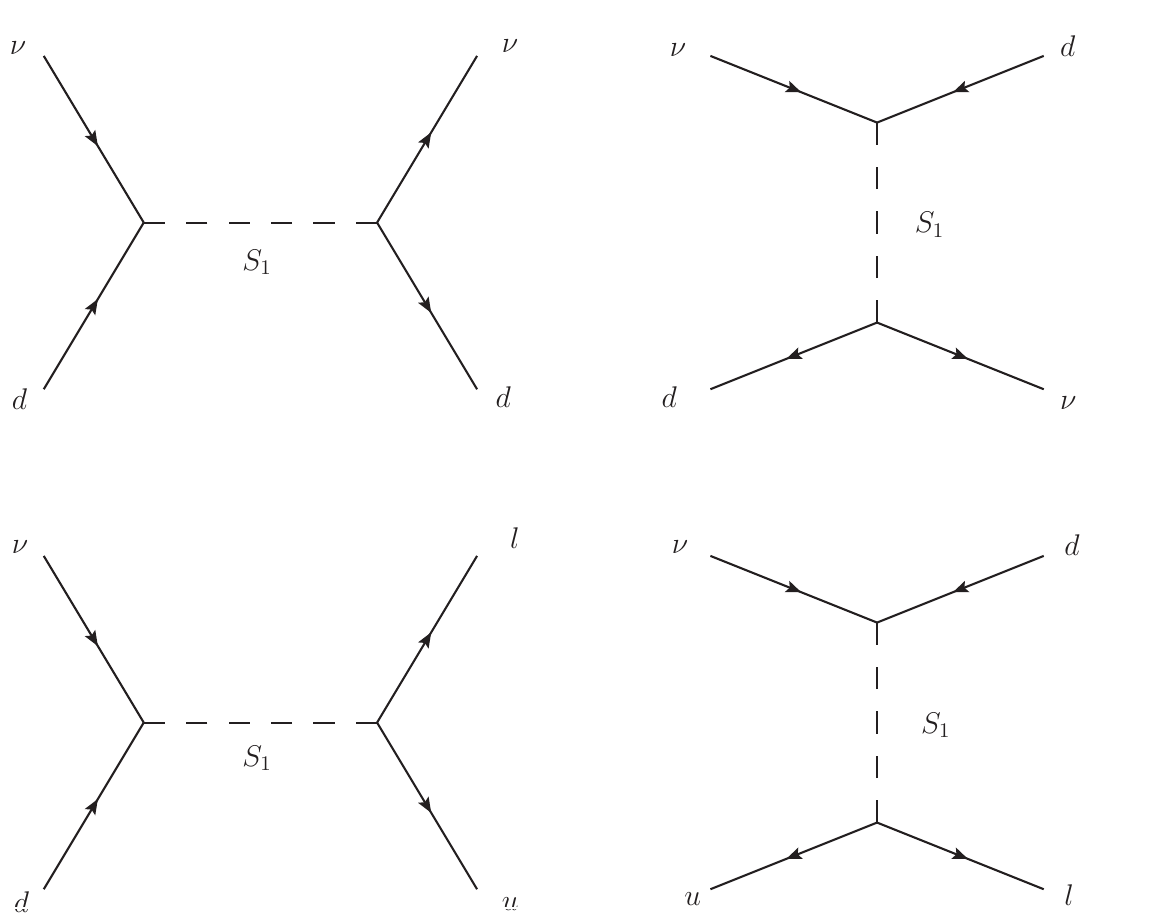}
\caption{Feynman diagrams for neutral current (upper diagrams) and charge current (down diagrams) in the neutrino-quark scattering process for the SM augmented with the LQ $S_1$. The arrows show the fermionic flow. }\label{fig:FeynmanDiag1} 
\end{figure}

In this section, we analyze the parameter space for the minimal SM augmented with the $S_1$ Leptoquark by using the restrictions found for the GNI couplings in section \ref{sec:results}. Besides, we compare with previous GNI restrictions coming from the CHARM and CDHS experiments~\cite{Escrihuela:2021mud}. The relevant terms of the Lagrangian in the gauge eigenstate basis can be written as 
\begin{equation}\label{eq:lagrangianLQ}
\mathcal L \supset y^L_{ij}\bar Q_i^c i \sigma_2 L_j S_1 + y^R_{ij}\bar u_i^c l_{jR} S_1 + \tilde y_{ij}^R \bar d_{iR}^c \nu_{jR} S_1 + \text{h.c},
\end{equation} 
where $Q^c_i$ and $L_j$ denote the left-handed quark and the lepton doublet with flavor indices $i,j$. The fields $u^c_R$ ($d_R$) and $l_R$ are the right-handed up-type (down-type) quark and charged lepton singlets, respectively. The superscript $c$ in the fermion fields stands for the charge conjugation field defined as $\Psi^c$, where 
\begin{equation}\label{eq:charge_conj}
\Psi^c = C \bar \Psi^T \quad  \text{and} \quad \bar \Psi^c = -\Psi^T C^{-1},
\end{equation}
with $C$ the charge conjugation matrix. The Yukawa coupling between $S_1$, charge conjugate quark, $q^C_i$, and lepton, $l_j$,  is denoted as $y_{ij}^{L,R}$, where $L,R$ stands for the lepton chirality. Note that $S_1$ couples to right-handed neutrinos with a Yukawa coupling $\tilde y^R_{i\nu_j}$, where we show explicitly the index $\nu_j$ to distinguish it from the right-handed LQ coupling to a quark-charged lepton pair. In our calculation, we assume all the LQ couplings to be real in order to simplify our analysis. The scalar LQ can mediate charge and neutral current interactions through the Feynman diagrams displayed in Fig~\ref{fig:FeynmanDiag1}. Since the LQ $S_1$ does not conserve leptonic and barionic number, the Feynman diagrams involve fermionic flows clashing or departing from the vertex, which require special treatment.

We have followed the approach of Refs.~\cite{Crivellin:2021ejk, Denner:1992vza}, which uses the properties of the charge-conjugation matrix $C$ to deduce the collection of Feynman rules for (anti-)fermionic external lines and vertices. In order to obtain the relations between the LQ couplings and the GNI parameters that appear in Eqs.~\eqref{LCC} and~\eqref{equation_1}, we write down the amplitudes for the process mediated by $S_1$ at first order in the $q^2/m^2_{S_1}$ expansion, where $q$ is the moment flowing through the LQ propagator and $m_{S_1}$ its mass. The LQ exchange yields to 4-fermion effective operators that must be Fierz arranged and identify with the operators of the effective Lagrangian. Besides, since the amplitude contains charge conjugated spinors, we use the relations in Eq.~\eqref{eq:charge_conj}, together with  $C \gamma_\mu = -\gamma_\mu^T C$ and $C \gamma_5 = \gamma^T_5 C$.

We introduce the phenomenological study for the charged and neutral currents separately.

\begin{itemize}
\item {\bf Charged current:} The scalar LQ can mediate the charged current processes $\nu_\alpha d \to \ell_\beta u$ and $\nu_\alpha \bar u \to \ell_\beta \bar d$. In this case, we obtain the following relations with the GNI parameters 

\begin{align}
\epsilon^{ud,L}_{\alpha   \beta} &= -\frac{1}{4\sqrt{2}G_F V_{ud}}\frac{y^L_{d\nu_\alpha}y^{L*}_{u \beta}}{m^2 _{S_1}},\label{eq:GNI_LQs} \\
\epsilon^{ud,S}_{\alpha \beta}   &= \frac{1}{4\sqrt{2}G_F V_{ud}}\frac{y^L_{d\nu_\alpha}y^{R*}_{u\beta}}{m^2_{S_1}}.
\end{align}

The pseudoscalar and tensor relationships are obtained by setting $ \epsilon^{ud,P}_{\alpha \beta} =  \epsilon^{ud,S}_{\alpha \beta}$ and $\epsilon^{ud,T}_{\alpha \beta} = -\epsilon^{ud,S}_{\alpha \beta}$. Then, we can calculate the sensitivities of the LQ coupling constants by utilizing the corresponding values for the GNI parameters listed in Tab~\ref{tab:limits}. These sensitivities are displayed in Tab~\ref{tab:Couplings1} as a function of $m_{S_1}$. As expected, the constraints on the parameters related to the tau neutrino flux are more relaxed compared to those imposed on the muon and electron neutrino flux.
 
\begin{table}[t]
\begin{center}
\begin{tabular}{c  c | c}
\hline  \hline
\multicolumn{2}{c}{LQ parameters}   & Sensitivity $\times (10^{-5}\: m^2_{S_1})$  \\ 
\hline \hline
\multirow{9}{*}{CC}  &$y^L_{d \nu_e} y^L_{u e}$      &$(6.332,7.082)\cup(-0.366,0.411)$\\
  &$|y^L_{d \nu_e} y^L_{u \beta}|$       &$<1.613 $\\
  &$|y^L_{d \nu_e} y^R_{u \beta}|$       &$<2.175 $ \\
  &$y^L_{d \nu_\mu} y^L_{u \mu}$        &$(6.381,7.103)\cup(-0.3525,0.3938)$\\
  &$|y^L_{d \nu_\mu} y^L_{u \beta}|$    &$<1.583$\\
  &$|y^L_{d \nu_\mu} y^R_{u \beta}|$    &$<2.139 $ \\
  &$y^L_{d \nu_\tau} y^L_{u \tau}$         &$ (-2.324,9.04)$\\
  &$|y^L_{d \nu_\tau} y^L_{u \beta}|$		&$< 4.586 $\\
  &$|y^L_{d \nu_\tau} y^R_{d u \beta}|$  & $<6.190$\\
\hline
\multirow{3}{*}{NC} &$(y^L_{d \nu_\mu})^2 $      &$	(-0.266,0.294)\cup(5.292,7.358)$ \\
  &$|y^L_{d \nu_\mu} y^L_{d \nu_\beta}|$      &$<1.248$ \\
  &$|y^L_{d \nu_\mu} y^R_{d \nu_\beta}|$     &$<1.336$  \\
\hline \hline
\end{tabular}
\caption{Expected $90 \%$ C.L. limits on different combinations of LQ couplings products for the analysis of Charge current and Neutral current neutrino scattering at \faser.}
\label{tab:Couplings1}
\end{center}
\end{table}

In the left panel of Fig.~\ref{fig:CC_Leptoquarks}, we present the expected sensitivity in the $m_{S_1}$ vs $y^L_{d\nu\alpha} y^L_{u \beta}$ plane for both electron and muon neutrinos in the cases of flavor-conserving (indicated by blue lines) and flavor-changing (indicated by red lines) interactions. It is worth mentioning that in the case of flavor conserving we only consider the intervals $(-0.366,0.411)$ for the electron neutrino and $(-0.3525,0.3938)$ for the muon neutrino. The right panel displays the sensitivity range for the $y^L_{d \nu_\alpha}$ and $y^L_{u \beta}$ couplings for muon neutrinos, assuming a fixed value of $m_{S_1}=1000$ GeV. In this scenario, we omit presenting the sensitivities of the electron and tau neutrinos. This is because the electron constraints are very similar to those of the muon neutrino, while the constraints on tau neutrinos are relatively less rigorous. Note that the  bounds for $y^L_{d\nu_\alpha}$ and $y^L_{u \ell}$ have been obtained by considering only the GNI parameter $\epsilon^{d,L}_{\alpha\beta}$  (Eq. \eqref{eq:GNI_LQs} ) different to zero.
\begin{figure}[hbt!]
\centering
\includegraphics[width=15cm]{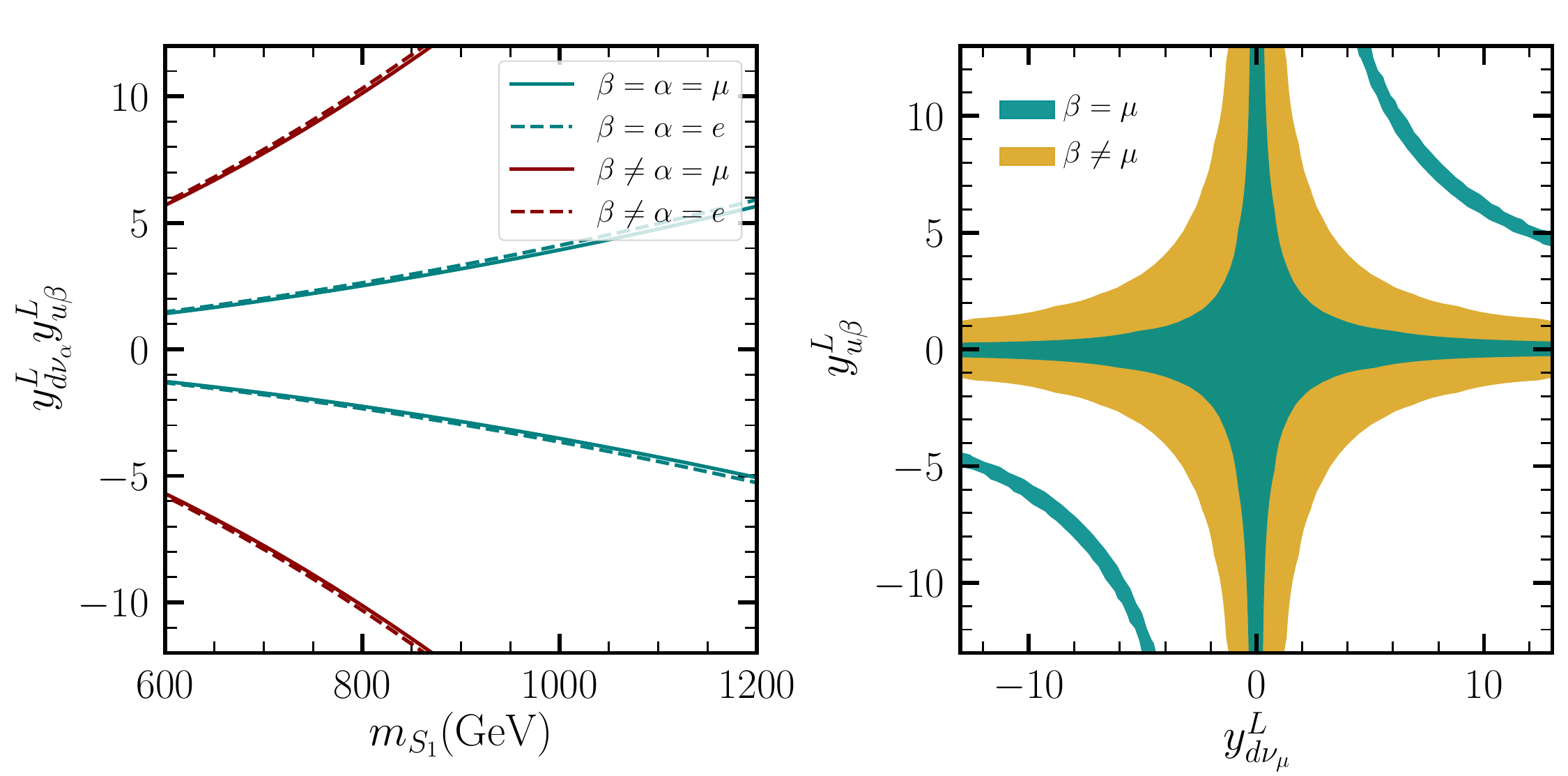}
\caption{Constraints on the simplified LQ model parameter space from the constraints on $\epsilon^{q,X}_{\alpha\beta}$ summarized in Table~\ref{tab:Couplings1}. The left plot shows the constrains over the product $y^{d\nu_\alpha}y^{L}_{u\beta}$, for $\alpha = \mu,e$ as function of $m_{S_1}$ in the cases of flavor conserving (blue lines) and flavor changing (red lines) interactions. The right plot shows the expected sensitivity in the $y^L_{d\nu_\mu}y^L_{u\beta}$ plane for $m_{S_1}=1000$ GeV.}\label{fig:CC_Leptoquarks} 
\end{figure}

By considering only the GNI parameters $\eps^{d,L}_{\alpha \beta}$ and $\epsilon^{d,T}_{\alpha\beta}$ as nonzero, we are able to determinate the allowed region in the $y^{L}_{u\mu}$ vs  $y^{R}_{u\mu}$ plane for $m_{S_1}=1000$ GeV. This region is depicted in Fig. \ref{fig:CC_LeptoquarksLR}, where we examine the scenarios in which the left handed coupling to neutrinos $y^L_{d\nu}$ takes the conservative values $1$ and $2$. 
\begin{figure}[hbt!]
\centering
\includegraphics[width=8.5cm]{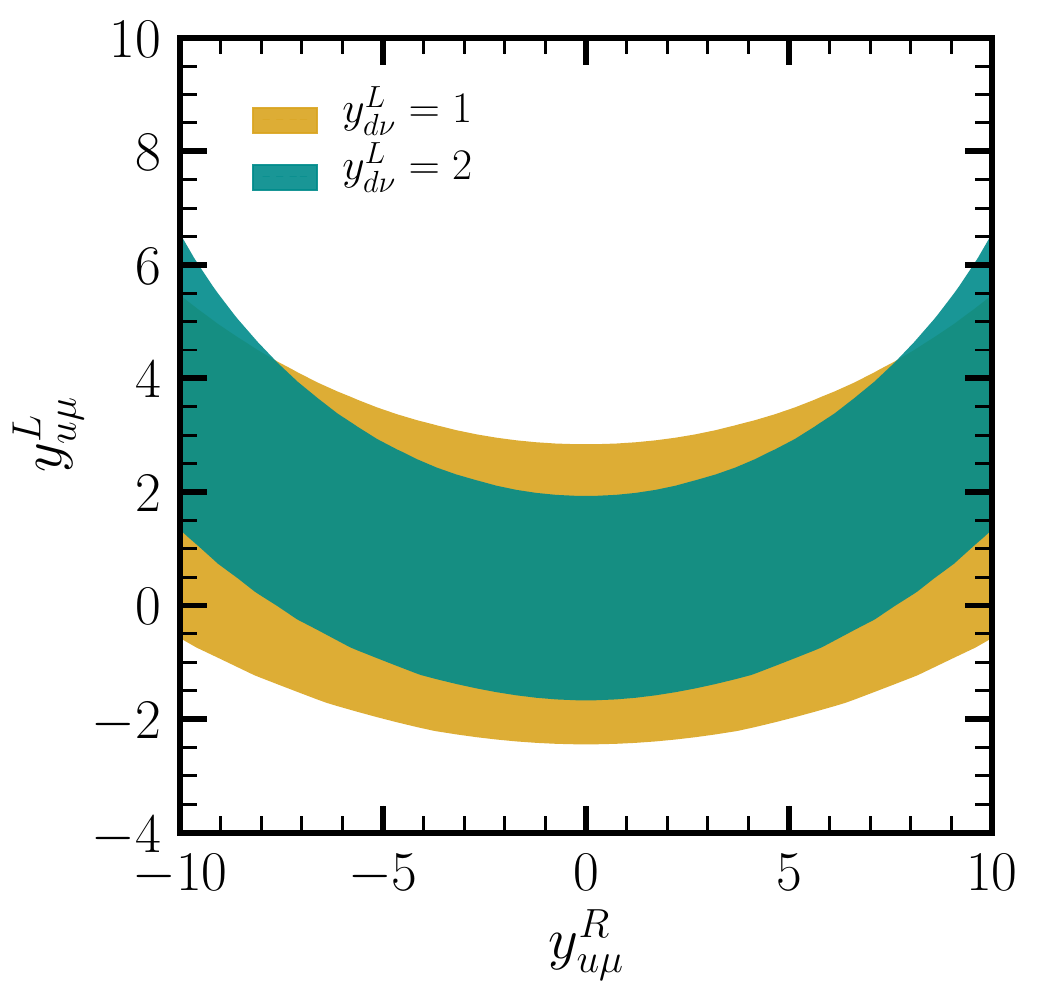}
\caption{Allowed area for the left and right LQ coupling to the quark up-$\mu$ pair for $m_{S_1}=800$ GeV and two values of the LQ coupling $y^L_{d\nu}$.}\label{fig:CC_LeptoquarksLR} 
\end{figure}
\item {\bf Neutral current:} The $S_1$ scalar Leptoquark can mediate the neutral current interactions $\nu_\alpha \overset{\scriptscriptstyle{(-)}}{d} \to \nu_\beta \overset{\scriptscriptstyle{(-)}}{ d}$, however, since $S_1$ does not have couplings with the up quark-neutrino pair, there is no contribution to the process $\nu \overset{\scriptscriptstyle{(-)}}{u} \to \nu \overset{\scriptscriptstyle{(-)}}{u}$. The following relationships between the parameters of LQ and GNI are obtained for this neutral current case
\begin{align}
\epsilon^{d,L}_{\alpha \beta}  &= \frac{1}{4 \sqrt{2} G_f}\frac{y^L_{d \nu_\alpha} y^{L*}_{d \nu_\beta}}{ m_{S_1}^2},\\
\epsilon^{d,S}_{\alpha \beta}  &= \frac{1}{4 \sqrt{2} G_f }\frac{y^L_{d \nu_\alpha} y^{R*}_{d \nu_\beta}}{m_{S_1}^2}. 
\end{align}
The relationships between the pseudoscalar and tensor GNI with the LQ parameters can be determined by setting $\epsilon^{d,P} = \epsilon^{d,S}$ and $\epsilon^{d,T} = -\epsilon^{d,S}/4$. Applying the constraints derived from the FASER data (presented in Tab.~\ref{tab:limits}), we use the aforementioned relationships to impose limitations on the $S_1$ couplings to fermions. The correspondings sensitivities, as a function of the Leptoquark mass, are shown in the lower part of Tab.~\ref{tab:Couplings1}. To illustrate, the left panel of Fig.~\ref{fig:NC_Leptoquarks} displays the constraints in the $y^L_{d\nu_\alpha}y^R_{d\nu_\beta}-m_{S_1}$ plane, while the right panel presents the allowed region in the $y^L_{d\nu_\alpha}-y^R_{d\nu_\beta}$ plane when the LQ mass is set to 1000 GeV. We also include the results obtained in Ref.~\cite{Escrihuela:2021mud}, where the authors carefully examined the GNI phenomenology using data from the CHARM and CDHS experiments. The green lines in this case represent the limitations on the corresponding LQ couplings when the combined analysis of data from CHARM and CDHS is performed. It is evident that the limits imposed by the CHARM and CDHS data are more stringent than the sensitivity expected by the \faser experiment. It is worth mentioning that if $y^L_{d\nu}=y^R_{d\nu}$, our findings are consistent with those provided in~\cite{Schwemberger:2023hee}, where the authors established limits on the $S_1$ LQ model by analyzing data from the LUX-ZEPLIN (LZ) experiment

\begin{figure}[hbt!]
\centering
\includegraphics[width=15cm]{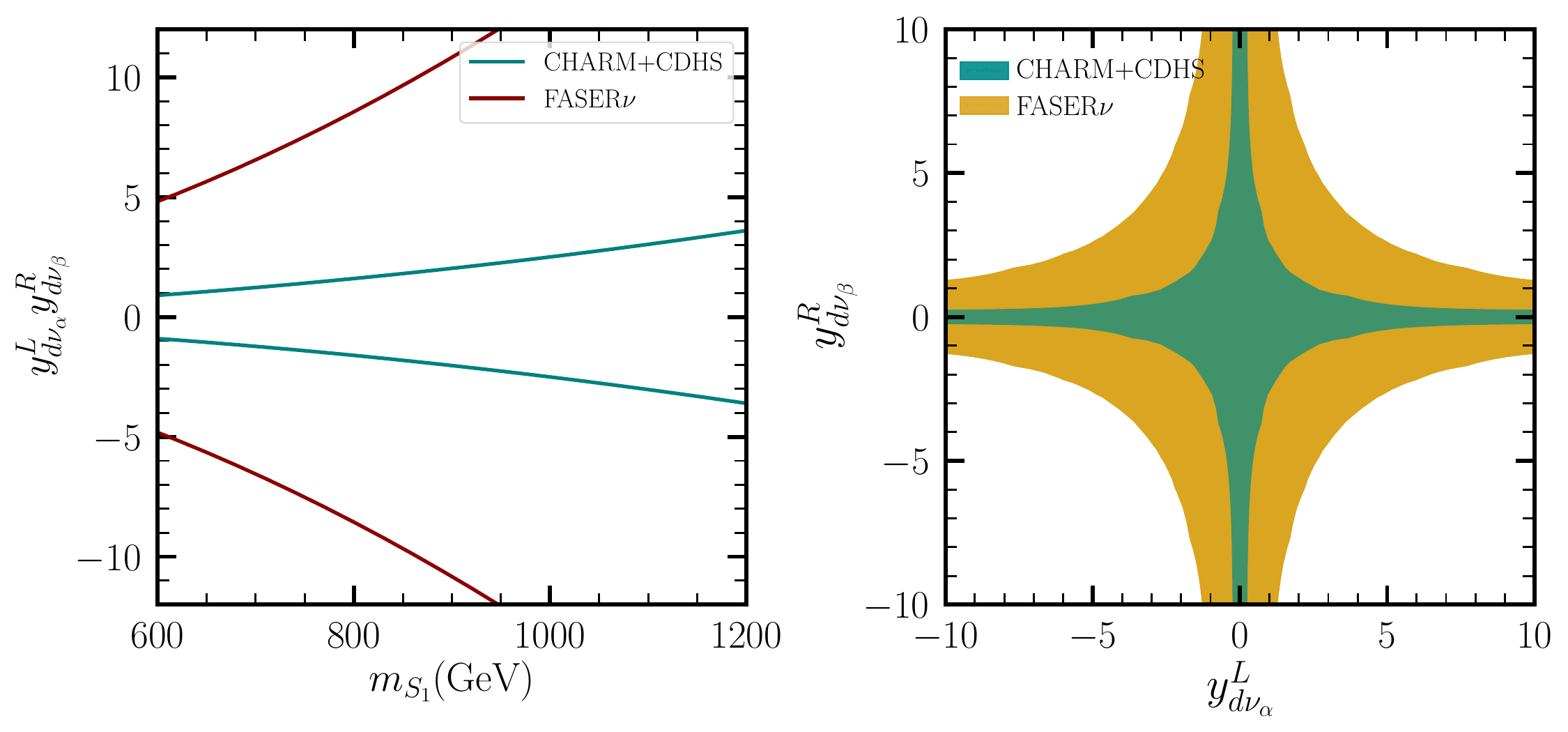}
\caption{Constraints on the simplified Leptoquark model parameter space from current constraints on $\epsilon^{f,j}_{\alpha \beta}$ summarized in table~\ref{tab:Couplings1}. The left plot shows the limits of the coupling product $y^L_{d\nu}y^R_{d\nu}$ as function of $m_{S_1}$, while the allowed areas in the plane $y^L_{d\nu}$ vs $y^R_{d\nu}$ for $m_{S_1}=1000$ GeV is displayed in the right plot.}\label{fig:NC_Leptoquarks} 
\end{figure}

\end{itemize}

\section{Conclusions}
\label{sec:conclusions}
In this work, we have studied the sensitivity of the FASER$\nu$ experiment to potentially constrain Generalized Neutrino Interactions. We have divided our analysis in two parts, one corresponding to the study of the charged current, and the other corresponding to the neutral current. We have set  bounds on all the different GNI parameters (scalar, pseudoscalar, vector, axial-vector and tensor) for both, charged and neutral interactions. First, regarding flavor, we see that the best constraints for the CC come from the parameters that couple with the muon neutrino, followed closely by those that couple with the electron neutrino. The parameters that couple with the tau neutrino are by far the weakest. Now, regarding the nature of the interaction, the stronger constraints for CC come from vector and axial-vector parameters, followed by the tensor parameters.\\
\indent For the NC case, focusing first on flavor, we see that the stronger constraints come from the parameters that couple with the down quark. On the other hand, focusing on the nature of the interaction, we have the same pattern that we had in the CC case, that is, the stringent constraints come from vector and axial-vector interactions, followed by those coming from tensor interactions. Scalar interactions are the least constrained. Despite FASER$\nu$ is not primarily thought as a neutral current experiment, following the work done in Ref.~\cite{Ismail:2020yqc}, but generalizing it to study GNI and not just NSI, we can see that the the potential of FASER$\nu$ to constraint NC is promising, and could be complementary with other experiments explicitly constructed to include the search for neutral current interactions. Just to make a comparison with other experiments, our bounds for NC parameters at FASER$\nu$ are of the same order of magnitude as the bounds we found in our previous work \cite{Escrihuela:2021mud} for the CHARM and CDHS experiments. In fact, the bounds for NC derived in this work for FASER$\nu$ are only a factor of 3 weaker than those we found there for CHARM and CDHS. 

Finally, we have employed a minimal Leptoquark model, denoted as $S_1$, to study the allowed parameter space by using the restrictions found in the analysis of the GNI couplings. Similar to the GNI case, we have divided the LQ analysis in CC and NC separately. For the CC scenario, we found restrictions over the Yukawa couplings $y^L_{d\nu_\alpha}$ and $y^R_{u\ell}$, while the NC analysis only allow us to derive constraints over the LQ couplings to neutrinos and down quarks $y^{R(L)}_{d\nu}$ . Inherited from the GNI restrictions, the LQ coupling to tau neutrinos has the weakest constraints, while the muon and electron neutrino couplings are very similar. The most restricted bounds are obtained for the LQ coupling product $y^L_{d\nu}y^L_{u\ell}$, which are derived for the vector and axial-vector GNI coupling. As for the NC case, we found limits on the LQ coupling product $y^L_{d\nu_\alpha}y^R_{d\nu_\beta}$ , where the muon neutrino constrains are favored over the electron and tau neutrinos. It turns out that in the minimal coupling scenario, we have found that the restrictions imposed by using the \faser data (this work) are very close by those using the LZ data.
\section{Acknowledgments}
This work has been partially supported by CONAHCyT research grant: A1-S-23238. The work of O. G. M., L. J. F., and R. S. V. has also been supported by SNI (Sistema Nacional de Investigadores, Mexico), and the work of J. R. has been supported by the program estancias posdoctorales por M\'exico of CONAHCyT.

\bibliographystyle{utphys} 
\bibliography{merged22}

\end{document}